\documentstyle[12pt]{article}
\def\ve{\vfil\eject}  
  \let\bsk=\bigskip
  
\let\qd=\quad \let\qqd=\qquad \def\qqqd{\qquad\qquad}

\let\a=\alpha \let\b=\beta \let\g=\gamma 
\let\e=\varepsilon   \let\th=\theta
  \let\l=\lambda 
  \let\p=\pi \let\r=\rho \let\s=\sigma
 \let\th=\theta \let\o=\omega \let\c=\chi 
\let\ph=\varphi   
   
  \let\D=\Delta

\def\0{\over } \def\1{\vec }     \def\2{{1\over2}} \def\4{{1\over4}}
\def\5{\bar }  \def\6{\partial } \def\7#1{{#1}\llap{/}}
\def\8#1{{\textstyle{#1}}}       \def\9#1{{\bf {#1}}}
 \def\llp{\hbox to 0pt{\hss /\hskip1.5pt}}
\def\llo{\hbox to 0.2pt{\hss /}} \def\llq{\hbox to 0pt{\hss /\hskip0.5pt}}
\def\so{\supset\hbox to 0pt{\hss $\displaystyle -$\hskip1pt}}

\def\<{\langle } \def\>{\rangle }  
 
\def \({\left( } \def \){\right) }

 \let\eq=\equiv    \let\aus=\in
  \let\ti=\tilde     \let\and=\wedge
\let\bi=\choose    
\def\|#1{{}_{\bigg|_{#1}}}

\def\mao#1{\mathop{\rm {#1}}\nolimits}  \def\tr{\mao{tr}} 
  \def\mod{\mao{mod}}
\def\gcd{\mao{gcd}}

\def\pmbf#1{\setbox0=\hbox{${#1}$}   \kern-.025em\copy0\kern-\wd0
      \kern.05em\copy0\kern-\wd0     \kern-.025em\raise.0433em\box0 }

\def\ZZ{\relax{\sf Z\kern-.3em Z}} \let\u=\cup
\def\ca{{\cal A}}  \def\cc{{\cal C}} 
\def\ce{{\cal E}}   \def\ch{{\cal H}}
 \def\co{{\cal O}} \def\cp{{\cal P}}
\def\i{{\rm i}}   \def\ch{\hat c}

\def\beq{\begin{equation}}    \def\eeq{\end{equation}}
\let\nn=\nonumber             \def\lleq#1{\label{#1}\eeq}
\def\beann{\begin{eqnarray*}} \def\eeann{\end{eqnarray*}}
\def\bea{\begin{eqnarray}}    \def\eea{\end{eqnarray}}
\def\lea#1{\label{#1}\eea}

{\begin{list}{}%
{\settowidth{\labelwidth}{#1}%
\setlength{\leftmargin}{\labelwidth}\addtolength{\leftmargin}{\labelsep}%
}}%
{\end{list}}%

\def\plb#1 #2 {Phys. Lett. {\bf#1B} #2 }
\def\phr#1 #2 {Phys. Rep. {\bf  #1} #2 } 
\def\npb#1 #2 {Nucl. Phys. {\bf B#1} #2 }
\def\aph#1 #2 {Ann. Phys. {\bf #1} #2 }  
\def\jmp#1 #2 {J. Math. Phys. {\bf #1} #2 }
\def\prd#1 #2 {Phys. Rev. {\bf D#1} #2 }
\def\prl#1 #2 {Phys. Rev. Lett. {\bf #1} #2 }
\def\rmp#1 #2 {Rev. Mod. Phys.  {\bf #1} #2 }
\def\zpc#1 #2 {Z. Phys. {\bf #1C} #2 }
\def\cmp#1 #2 {Comm. Math. Phys. {\bf #1} #2 }
\def\mpl#1 #2 {Mod. Phys. Lett. {\bf A#1} #2 }

\def\eff{_{_{\rm eff}}}  \def\tt#1{\ti\th_{#1}}
\def \PP{Poincar\`e polynomial }

\def\inbar{\vrule height1.5ex width.4pt depth0pt}
\def\IB{\relax{\rm I\kern-.18em B}}
\def\IC{\relax\,\hbox{$\inbar\kern-.3em{\rm C}$}}
\def\ID{\relax{\rm I\kern-.18em D}}
\def\IE{\relax{\rm I\kern-.18em E}}
\def\IF{\relax{\rm I\kern-.18em F}}
\def\IG{\relax\,\hbox{$\inbar\kern-.3em{\rm G}$}}
\def\IH{\relax{\rm I\kern-.18em H}}
\def\II{\relax{\rm I\kern-.18em I}}
\def\IK{\relax{\rm I\kern-.18em K}}
\def\IL{\relax{\rm I\kern-.18em L}}
\def\IM{\relax{\rm I\kern-.18em M}}
\def\IN{\relax{\rm I\kern-.18em N}}
\def\IO{\relax\,\hbox{$\inbar\kern-.3em{\rm O}$}}
\def\IP{\relax{\rm I\kern-.18em P}}
\def\IQ{\relax\,\hbox{$\inbar\kern-.3em{\rm Q}$}}
\def\IR{\relax{\rm I\kern-.18em R}}
\def\ZZ{\relax{\sf Z\kern-.4em Z}}
\def\beq{\begin{equation}}  \def\eeq{\end{equation}}
\def\bea{\begin{eqnarray}}  \def\eea{\end{eqnarray}}
\def\notin{\ \hbox{{$\in$}\kern-.51em\hbox{/}}}
\def\nn{\nonumber}
\def\tabroom{\hbox to0pt{\phantom{\Huge A}\hss}}
\def\a{\alpha}  \def\b{\beta}   \def\e{\epsilon} \def\g{\gamma}
  \def\l{\lambda} \def\s{\sigma}   \def\th{\theta}
\def\bn{\bar n} \def\bz{\bar z} \def\bth{\bar {\theta}}

  \def\cL{{\cal L}} 
   
 \def\cT{{\cal T}} 

\def\notin{\ \hbox{{$\in$}\kern-.51em\hbox{/}}}
\def\fnote#1#2{\begingroup\def\thefootnote{#1}\footnote{#2}\addtocounter
{footnote}{-1}\endgroup}
\def\tabroom{\hbox to0pt{\phantom{\Huge A}\hss}}
\let\ce=\centerline

\begin{document}
{\hfill NSF--ITP--91--102}\vskip -10pt
{\hfill TUW--91--10}      \vskip -10pt
{\hfill (corrected)}
\vskip 15mm
\centerline{\large ABELIAN LANDAU--GINZBURG ORBIFOLDS AND MIRROR SYMMETRY}

\vskip 15mm
\ce{\sc M. Kreuzer$^*$ \fnote{\sharp}{Present address:
			 CERN, TH-Division, CH-1211 Gen\`eve 23, Switzerland},~
	R. Schimmrigk$^{\diamondsuit}$
	\fnote{\natural}{Present address: Institut
	       f\"ur Theoretische Physik, Universit\"at
	       Heidelberg, 6900 Heidelberg, Germany}
	\fnote{\bullet}{Supported in part by NSF Grant No. PHY 89--04035}
	and H. Skarke$^*$}
\vskip .5truein
\ce{$^*$ \it Institut f\"ur Theoretische Physik, Technische Universit\"at Wien}
\ce{\it  Wiedner Hauptstrasse 8--10, 1040 Wien}
\ce{\it  AUSTRIA}
\vskip .3truein
\ce{$^{\diamondsuit}$\it Institute for Theoretical Physics,
                         University of California}
\ce{\it  Santa Barbara, CA 93106}
\ce{\it  USA}
\vfill 
\ce{\bf ABSTRACT}\bsk

We construct a class of Heterotic String vacua described by
Landau--Ginzburg theories  and consider orbifolds of these models
with respect to abelian symmetries. For LG--vacua described by potentials
in which at most three scaling fields are coupled we explicitly construct
the chiral ring and discuss its diagonalization with respect to its most
general abelian symmetry. For theories with couplings between at most
two fields we present results of an explicit construction of the LG--potentials
and their orbifolds. The emerging space of (2,2)--theories shows a remarkable
mirror symmetry. It also contains a number of new three--generation models.

\renewcommand\thepage{}
\vfill
\eject

\baselineskip=17pt 
\pagenumbering{arabic}

\section{Introduction}

The question of the structure of the configuration space of string theory
is an important one for theoretical as well as practical reasons.
Unfortunately
there are few well developed tools available for a general analysis of this
space from first principles. Instead much of the insight gained over the
last years stems from explicit constructions of string vacua. These
use a variety of methods ranging from lattice techniques
and exactly solvable models to mean field theory and algebraic geometry.

Recently techniques from Landau--Ginzburg mean field theory have been
utilized to construct a set of several thousand consistent Heterotic
vacua \cite{cls}. This construction extends the number of known vacua
by an order of magnitude
\footnote{Here we consider those vacua as distinct that have a different
          number
          of generations or antigenerations; this is a very rough measure
          since it does not take into account the Yukawa couplings between
	  these fields.}
and provides a large enough slice of the moduli
space to expose an important property of this space, its {\it mirror
symmetry}. Considering that these models have been constructed as
completely independent LG theories this provides strong evidence that the
space of left--right symmetric Heterotic String vacua indeed features
mirror symmetry.

An a priori independent technique of constructing Heterotic vacua was
pursued in refs. \cite{gp}\cite{fkss}\cite{fiqs}. The starting point of
those
papers is the set of exactly solvable N=2 superconformal tensor models
\cite{g} constructed explicitly in \cite{fkss} \cite{ls1}. These models
always have
discrete symmetries and hence it is possible to consider orbifolds of
any of these tensor models by modding out any of the subgroups of
their symmetries. It was observed in refs. \cite{fiqs}\cite{fkss}\cite{gp}
that in some cases this orbifolding procedure produces mirror pairs
\footnote{An analysis of mirror orbifolds of Calabi--Yau manifolds of
          Fermat type in weighted $\IP_4$ has been performed in
	  \cite{r}.}.

It turns out that these two modes of construction are not completely
independent. It was shown in ref. \cite{ls2} that certain classes of mirror
pairs of vacua can be related via a process
involving two steps: first a LG--vacuum is orbifolded and then the
order parameters of the LG--potential are transformed into new
fields with a nonlinear transformation involving fractional powers.
This technique can be  applied not only to mean field theories associated
to the exactly solvable models but also to the much more general class
constructed in \cite{cls}.
Its application is not restricted to mirror pairs but is completely
general, depending only on the type of symmetry considered. Hence this
result suggests that a general relation might exist
between Landau--Ginzburg potentials and their orbifolds. To
investigate this question further it is clearly useful to consider
orbifolds of the tensor models and, more generally, LG theories in a
systematic way.

Our results indeed show that there is substantial overlap between the
Landau--Ginzburg theories constructed in [1] and our orbifolds. Figure 1
shows a plot of the difference between the number of generations and
antigenerations versus the sum of these numbers for the class of LG theories
we constructed and all their orbifolds with respect to phase symmetries.
Similar to the results in [1] the diagram shows a remarkable symmetry
with respect to the exchange $n_g\longleftrightarrow \bar n_g$. Even
though our implementation is not complete already $94\%$ of
the Hodge pairs have mirror partners.


A second motivation for our work is the fact that it is surprisingly
difficult to find ground states of the Heterotic String theory that
accommodate the Standard Model in a painless manner.
Despite all interest in the general structure of the configuration
space the search for realistic models remains an important challenge.
Knowledge of the general structure is, after all, aimed toward a mechanism
to lift the degeneracy of the groundstates and hence a much more ambitious
goal.

In the present paper we construct a class of Landau--Ginzburg theories
which contains the class of minimal tensor models as a small subset.
We work out the formulae needed for the evaluation of the
number of 27 and ${\overline {27}}$ $E_6$--representations
for any abelian orbifold of a large set  of LG--vacua.
Then we proceed to orbifold them with respect to abelian
symmetries with determinant 1.
These theories correspond to supersymmetric orbifolds of Calabi--Yau
spaces.

This paper is organised as follows:
In section 2 we briefly review the results of Vafa and Intriligator
\cite{v} \cite{iv} on the construction
of LG orbifolds. In section 3 we determine the local algebra, which
corresponds to the chiral ring of the SCFT, for a class of quasihomogeneous
singularities. In section 4 we define the class of models { which we
considered in our explicit constructions.}
In section 5 the diagonalization of the chiral ring with
respect to its most general discrete abelian symmetry  and the eigenvalues and
dimensions of the eigenspaces are computed.
Section 6 contains some general considerations
about the symmetries we have implemented. In section 7 we present our
results for phase symmetries and in section 8 for cyclic symmetries.
Finally we present our conclusions.

\section{Landau -- Ginzburg orbifolds}
The Landau--Ginzburg description of an $N=2$ SCFT \cite{lvw} is
determined by an action of the form
\beq \int d^2zd^4\th K(\Phi_i,\bar\Phi_i)+
     \left(\int d^2zd^2\th W(\Phi_i)+c.c.\right),\eeq
where the superpotential $W$ is quasihomogeneous of degree $d$ in
the chiral superfields $\Phi_i(z,\bz,\th^{\pm},\bth^{\pm})$
of weight $k_i$
\beq  W(\l^{k_i}\Phi_i)=\l^d W(\Phi_i)                                    \eeq
with an isolated singularity at $\Phi_i=0$. The central charge of the
superconformal theory is given by the highest weight
\beq \qqd\hat c={c\03}=\sum_i(1-2q_i) \lleq{lgp}
with $q_i={k_i\0d}$. Its chiral ring is isomorphic to what mathematicians
call the local algebra of the nondegenerate quasihomogeneous function $W(z_i)$,
defined as the ring of all polynomials in some complex variables $z_i$
modulo the ideal generated by $dW/dz_i$ \cite{agv}.
In the present context the $z_i$ denote the constant values of the
lowest components of the chiral superfields. The zero locus of $W$
defines a complex variety in some complex space $\IC_n$.
We will use a short hand and denote the space of such polynomials by
\beq
\IC_{(k_1,k_2,\dots ,k_n )}[d]  \lleq{aff}
and call it a configuration. For the computation of the spectrum of the LG
theory it is not important to know the precise form of the polynomial; only
the set of weights $k_i$ is important as well as the fact that the
configuration does have a member with an isolated singularity.
The nondegeneracy of $W$ implies that the local algebra
(and therefore the chiral ring) is finite dimensional. The Poincar\`e
polynomial $P(t)$ is defined as the generating function for the number of
basis monomials of the local algebra of a specific degree of quasihomogeneity,
i.e. the number of states of a given conformal weight. It can be computed
with the formula
\beq P(t)=\prod(1-t^{1-q_i})/(1-t^{q_i}). \lleq{pp}
Note that this expression is not a polynomial in
$t$ but rather in $t^{1/d}$.
For convenience, however, we will refer to $P(t)$ and not to $P(t^d)$ as the
Poincar\`e polynomial.

If we want to use an N=2 superconformal theory for constructing a
string vacuum with N=1 spacetime supersymmetry, we require $\hat c=3$
and integral $U(1)$ charges. By orbifolding an LG theory it is possible
to obtain a theory with integral charges.
One way to achieve this is to orbifoldize the theory with respect
to the U(1) symmetry of the $N=2$ superconformal algebra. Since we are
considering only rational theories all the fields have rational charges
and hence this U(1) projection translates, in the mean field description
of the superconformal theory, into an orbifolding  with respect to the
$\ZZ_d$ symmetry of the superpotential $W(\Phi_i)$.
In this case the numbers of states with charges $(q_L,q_R)$ are
given \cite{v} by the coefficients of $t^{q_L}\bar t^{q_R}$ in
\beq P(t,\5t)=\tr\,t^{J_0}\5t^{\5J_0}
   =\sum_{0\le l<d}\prod_{\tt i\aus\ZZ}{1-(t\5t)^{1-q_i}\01-(t\5t)^{q_i}}
   \prod_{\tt i\not\aus\ZZ}(t\5t)^{\2-q_i}\({t\0\5t}\)^{\tt i-\2}\|{int}, \eeq
where $\tt i=\th_i-[\th_i]$ and $\th_i=lq_i$ is the non-integer part of
$lq_i$. The subscript $int$ means that only integral powers of $t$ and
$\bar t$ are kept in this expression. If we have a Calabi--Yau interpretation
of our theory, these coefficients correspond to the Hodge numbers of the
CY--manifold. For $\hat c=3$
\beq P(t,\bar t)=(1+t^3)(1+\5t^3)+n_g(t\5t+t^2\5t^2)+\5n_g(t\5t^2+t^2\5t),
                                                                       \eeq
where $n_g$ and $\bar n_g$ denote the numbers of $27$ and ${\overline {27}}$
representations of $E_6$ occurring in the construction of the Heterotic
String vacua. If it is possible to give masses to all $(27,{\overline {27}})$
pairs, then the Euler number $\c=2(\5n_g-n_g)$ is twice the net
number of fermion generations.

Of course, if our potential $W$ has more symmetries we are free to
orbifoldize with respect to any of them, and if certain constraints are
imposed then a consistent vacuum is obtained \cite{iv}. The formula
given above has to be
modified because, unfortunately, the Poincar\`e polynomial does not
contain the information
on the transformation properties of the states under general symmetries.
So for genuine orbifolds we need to rely on an explicit basis of the
chiral ring.
The expression for the
left and right charges of a state $\prod X_i^{\l_i}|0\>_{NS}$ in some
twisted sector, however, remains valid,
\beq q_\pm=\sum_{\tt i>0}(\8\2-q_i\pm(\tt i -\8\2))+\sum_{\tt i=0}\l_iq_i
                                                               \lleq{qpm}
with $\th_i$ now being the phase of the $i^{th}$ field in a diagonal basis
under the action of the group element defining the twist.
Note that in general the twisted vacua have nontrivial transformation
properties  under all symmetries (see \cite{iv}).

\section{Local algebra of quasihomogeneous functions}

In this section we work out the local algebra for all
nondegenerate quasihomogeneous functions of three or less complex variables.
Since our explicit  construction of potentials with $c=9$ and the
implemention of
symmetries discussed in later sections covers only models with couplings of
up to two superfields the reader who is more interested in the results
for the emerging string vacua rather than the general theory may wish to
skip some of the details in the present section.

The nondegenerate quasihomogeneous functions of three or less complex variables
have been classified in the mathematical literature \cite{agv}.
They are sums of functions of the form
\bea (\rm{I})\qquad &&z_1^{a_1}z_2+\cdots z_{n-1}^{a_{n-1}}z_n+z_n^{a_n},
                                                               \label{chain}\\
     (\rm{II})\qquad &&z_1^{a_1}z_2+\cdots z_{n-1}^{a_{n-1}}z_n+z_n^{a_n}z_1,\\
     (\rm{III})\qquad &&z_1^az_2+z_2^b+z_2z_3^c+\e z_1^pz_3^q,\\
   (\rm{IV})\qquad &&z_1^az_2+z_2^bz_3+z_2z_3^c+\e z_1^pz_3^q, \label{iiiv}\eea
where the first two types represent nondegenerate quasihomogeneous functions
for any $n\ge 1$ or 2.

The variables in a {\bf Type (I)} function have degrees of quasihomogeneity
\beq q_i ={1\0 a_i}-{1\0 a_ia_{i+1}}+\cdots +(-1)^{n-i}
      {1\0 a_ia_{i+1}\cdots a_n}.                                    \eeq
Using the abbreviations $x_i:=t^{q_i}$, the Poincar\`e polynomial fulfils
the recursion relation
\beq P_n(t;a_1,\cdots,a_n)={x_1^{a_1-1}-1\0 x_1-1}{x_2^{a_2}-1\0 x_2-1}\cdots
     {x_n^{a_n}-1\0 x_n-1}+x_1^{a_1-1}P_{n-2}(t;a_3,\cdots,a_n)           \eeq
with $P_1(t;a)=(x^a-1)/(x-1)$ and $P_0(t)=1$.
The local algebra is determined by the equations
\beq a_1 z_1^{a_1-1}z_2=z_1^{a_1}+a_2z_2^{a_2-1}z_3=\cdots=z_{n-1}^{a_{n-1}}
     +a_nz_n^{a_n-1}=0.                                               \eeq
The monomials $\prod_iz_i^{\a_i}$ with $\a_1\le a_1-2$ and $\a_i\le a_i-1$
for all other $i$'s are nonvanishing and independent, and all other monomials
with $\a_1\le a_1-2$ or $\a_1\ge a_1$ can be written as linear combinations
of them. Monomials with $\a_1 = a_1-1$ can only be nonvanishing if $\a_2=0$.
They are independent if and only if the other exponents correspond to a chiral
ring of the same type for $z_3,\cdots,z_n$, as the form of the \PP suggests.
The highest weight is $\ch=n-2\sum q_i$ with
\beq \sum_{i=1}^n q_i=\sum_{i=1}^n{1\0 a_i}-\sum_{i=1}^{n-1}{1\0 a_ia_{i+1}}
     +\sum_{i=1}^{n-2}{1\0 a_ia_{i+1}a_{i+2}}+\cdots +
     (-1)^{n-1}{1\0 a_1a_2\cdots a_n}.                                  \eeq

For {\bf Type (II)} the degrees are given by
\beq q_i(1+(-1)^{n-1}a_1\cdots a_n) =1-a_{i-1}+a_{i-1}a_{i-2}-
     a_{i-1}a_{i-2}a_{i-3}+\cdots +(-1)^{n-1}a_{i-1}a_{i-2}\cdots a_{i+1},
                                                                        \eeq
where the indices are to be understood modulo $n$.
With the same abbreviations as before, the Poincar\`e polynomial is
\beq P(t)={x_1^{a_1}-1\0 x_1-1}{x_2^{a_2}-1\0 x_2-1}\cdots{x_n^{a_n}-1\0
     x_n-1}.                                                            \eeq
The local algebra is determined by the equations
\beq z_{i-1}^{a_{i-1}}+a_iz_i^{a_i-1}z_{i+1}=0.                         \eeq
The monomials $\prod_iz_i^{\a_i}$ with $\a_i\le a_i-1$ form a basis.

The analysis for the types (III) and (IV), for which not all coefficients
can be normalized to 1, is more complicated.

The degrees in {\bf Type (III)} are
\beq q_1={1\0 a}\left(1-{1\0 b}\right),\quad q_2={1\0b},
     \quad q_3={1\0 c}\left(1-{1\0 b}\right),\eeq
yielding
\beq \sum 1-2q_i=3-2{ac+(b-1)(a+c)\0 abc}\\
     =(b-1){2ac+(p-2)c+(q-2)a\0 abc}  \eeq
for the highest weight. Quasihomogeneity of $z_1^pz_3^q$ implies
\beq \({p\0a}+{q\0 c}\)\(1-{1\0 b}\)=1,\quad {p\0a}+{q\0 c}=1+{1\0 b-1}.
                                                                  \lleq{pq}
Therefore, in order to allow a suitable last term for given values of $a$
and $c$, $b-1$ must be a divisor of the least common multiple of $a$ and
$c$. This is exactly the condition for the expression (\ref{pp})
for the \PP to be a polynomial.
With $x_1=t^{c(b-1)},\, x_2=t^{ac},\, x_3=t^{a(b-1)}$ and
$T=x_1^a=x_3^c=t^{ac(b-1)}$, one can show that the \PP is determined
by
\bea P(t)\simeq{x_1^{a-1}-1\0 x_1-1}{x_2^{b-1}-1\0 x_2-1}{x_3^{c-1}-1\0 x_3-1}
     +{x_1^{a-1}-1\0 x_1-1}(1+T+T^2)x_3^{c-1}\nn \\ +{x_3^{c-1}-1\0 x_3-1}
     (1+T+T^2)x_1^{a-1} +(1+T)x_1^{a-1}x_3^{c-1}\nn \\+{x_1^{a-1}-1\0 x_1-1}
     {x_3^{c-1}-1\0 x_3-1}(2T+T^2+T^3-Tx_1^{p-1}x_3^{q-1}),         \lea{piii}
where $\simeq$ stands for ``equality up to the highest weight'', i. e. on
the r.h.s. we neglect powers of $t$ that are higher than in the highest weight
term $T^2 x_1^{p-2}x_3^{q-2}$. The local algebra is determined by the equations
\beq az_1^{a-1}z_2+\e pz_1^{p-1}z_3^q=z_1^a+bz_2^{b-1}+z_3^c=cz_2z_3^{c-1}+
     \e qz_1^pz_3^{q-1}=0.                         \eeq
We construct a basis of the local algebra in the following way: The first
step is ``$z_2$ elimination'': While possible, we use the second and then
the first and third equation in order to substitute monomials by polynomials
which contain smaller powers of $z_2$. The remaining $z_2$--dependent monomials
are of the form
$z_1^\a z_2^\b z_3^\g$ with $0\le \a \le a-2, 1\le \b\le b-2$ and
$0\le\g\le c-2$. They are obviously represented by the first term in our
expression for the \PP. All other basis monomials can be chosen of the form
$z_1^\a z_3^\g$. Eliminating $z_2$ in a combination of the first and third
equation, we obtain
\beq cpz_1^{p-1}z_3^{c+q-1}=aqz_1^{a+p-1}z_3^{q-1}.               \lleq{oeeq}
This restriction on the independence of different monomials is related to
the last term in $eq.$ (\ref{piii}). Further equations are found by
combining the first or last two equations:
\bea z_1^{2a-1}+z_1^{a-1}z_3^c&=&a^{-1}\e bpz_1^{p-1}z_2^{b-2}z_3^q,
            \label{res2}\\
     z_1^az_3^{c-1}+z_3^{2c-1}&=&c^{-1}\e bqz_1^pz_2^{b-2}z_3^{q-1},
                                                         \label{res3}   \eea
where the r.h.s.'s are still subject to ``$z_2$ elimination''.
The arbitrariness of $\e$ permits us to neglect terms proportional to
$\e$ in the explicit construction of some basis of the algebra,
once $eq.$ (\ref{oeeq}) is derived. One thus  shows that a completion
of the basis is given by $z_1^\a z_3^\g$ with
$(\a, \g)\in [0,a-2]\times[0,2c-2]\cup[a-1,a+p-2]\times[0,c-1]\cup
 [a+p-1,2a+p-2]\times[0,q-2]$ for $p\ge a$ or $q\ge c$, and with
$(\a, \g)\in [0,p-2]\times[0,2c-2]\cup[p-1,a+p-2]\times[0,c+q-2]\cup
[a+p-1,2a-2]\times[0,c-2]$ otherwise. This is true for all $\e\in\IC$ except
for a finite number of values for which the singularity is degenerate.
Note that equations (\ref{res2}) and (\ref{res3}) are not completely
independent. Multiplying the first of them
with $z_3^{c-1}$ yields, after $z_2$ elimination, the same as multiplying the
second one with $z_1^{a-1}$. Even taking this into account, when we go to
high degrees, we have more restrictions than there are monomials.
If these restrictions were not redundant, the
$x_2$-independent part of the \PP would be (exactly)
\beq {1\0 1-x_1}{1\0 1-x_3}(1-Tx_1^{a-1}-Tx_3^{c-1}+Tx_1^{a-1}x_3^{c-1}
     -Tx_1^{p-1}x_3^{q-1}),                                       \lleq{ppp}
with $x_1$ and $x_3$ interpreted as counting $z_1$ and $z_3$, respectively.
This expression is equal to the one given above except for terms
exceeding the highest weight.
We conclude that the interpretation of the \PP in the form (\ref{piii}) or
(\ref{ppp})
is unique up to $T=x_1^a=x_3^c=t^{ac(b-1)}$; $x_1$ is interpreted as the
formal variable for counting $z_1$, $x_3$ is interpreted as counting $z_3$.

The analysis of {\bf Type (IV)} is similar to the previous case.
The degrees are
\beq q_1={c\0 a}{b-1\0 bc-1},\quad q_2={c-1\0 bc-1},\quad
     q_3={b-1\0 bc-1},                                            \eeq
yielding the highest weight
\beq \sum 1-2q_i=3-2{ab+ac+bc-2a-c\0 a(bc-1)}=
                  (b-1){2ac+(p-2)c+(q-2)a\0 a(bc-1)}.                 \eeq
Quasihomogeneity of $z_1^pz_3^q$ implies
\beq \({pc\0a}+q\){b-1\0 bc-1}=1,\quad {p\0a}+{q\0 c}=1+{c-1\0 c(b-1)}. \eeq
Thus for given $a$ and $c$ $b-1$ has to be a divisor of ${\rm lcm}(a,c)\cdot
(c-1)/c$. With $x_1=t^{c(b-1)},\, x_2=t^{a(c-1)},\, x_3=t^{a(b-1)}$ and
$T=x_1^a=x_3^c=t^{ac(b-1)}$ the \PP is given by
\bea P(t)\simeq{x_1^{a-1}-1\0 x_1-1}{x_2^{b-1}-1\0 x_2-1}{x_3^c-1\0 x_3-1}
     +x_2^{b-1}{x_1^{a-1}-1\0 x_1-1}+{x_1^{a-1}-1\0 x_1-1}(T+T^2)x_3^{c-1}
     \nn\\ +{x_3^{c-1}-1\0
x_3-1}(1+T+T^2)x_1^{a-1}+(1+T)x_1^{a-1}x_3^{c-1}\nn\\
     +{x_1^{a-1}-1\0 x_1-1}{x_3^{c-1}-1\0 x_3-1}
    (2T+T^2+T^3-Tx_1^{p-1}x_3^{q-1}).                           \eea
The local algebra is determined by the equations
\beq az_1^{a-1}z_2+\e pz_1^{p-1}z_3^q=z_1^a+bz_2^{b-1}z_3+z_3^c=
     z_2^b+cz_2z_3^{c-1}+\e qz_1^pz_3^{q-1}=0.                         \eeq
Again the first step in the construction of a basis of the local algebra
is ``$z_2$ elimination''. We end up with $z_2$-dependent monomials
of the form
$z_1^\a z_2^\b z_3^\g$ with $0\le \a \le a-2, 1\le \b\le b-2$ and
$0\le\g\le c-1$ or with $0\le \a \le a-2, \b=b-1$ and $\g=0$.
They are represented by the first two terms in our
expression for the \PP. With arguments similar to the ones used before,
one can show that the
part containing only $z_1$ and $z_3$ is the same as in the previous model.

\section{Fermats, loops, and tadpoles }

As already mentioned we have in our explicit construction of
LG--potentials and their orbifolds focused on models described by
superpositions of potentials in which at most two fields interact.
 In this section we thus specialize the above
considerations to this case and describe how we obtained our list of 7579
different LG potentials.

We call a contribution $\Phi^a$ to the potential Fermat type,
whereas the simplest possible couplings look like
\bea W^{(L)}=\Phi^e \Psi+\Phi \Psi^f, \qqd q_\Phi={f-1\0ef-1},&\qd&
             q_\Psi={e-1\0ef-1},\qd  \ch^{(L)}=2{(e-1)(f-1)\0ef-1},\\
     W^{(T)}=\Phi^e +\Phi \Psi^f, \qqd \qqd q_\Phi={1\0e},&\qd&
                        q_\Psi={e-1\0ef},\qqd  \ch^{(T)}=2{(e-1)(f-1)\0ef}\eea
and are called loops and tadpoles, respectively.

The corresponding chiral rings are represented by linear combinations
of monomials of the form $\Phi^i$, $0\le i\le a-2$ for Fermat type. A basis
for loops and tadpoles is given by
\bea L: \{\Phi^i\Psi^j,&\;& 0\le i\le e-1,\; 0\le j\le f-1\},\label{L}\\
     T: \{\Phi^i\Psi^j,&\;& 0\le i\le e-1,\; 0\le j\le f-2\}
        \;\cup\;\{\Psi^{f-1}\}.                 \label{T} \eea
These formulae are of course contained in the results for type (I) and (II) in
the previous section.

It is now straightforward to construct all possible combinations of these
potentials which add up to a total central charge of $c\eq3\ch=9$. First we
note that the accumulation points of possible contributions $\ch$ are
$2{i-1\0i}$ with $2\le i\le\infty$. If we define the finite set
\beq \textstyle M_\e=\{\ch\}-(1-\e,1)\u({4\03}-\e,{4\03})\u({3\02}
     -\e,{3\02})\u({8\05}-\e,{8\05})\u({5\03}-\e,{5\03})\u({12\07}-\e,2),\eeq
one can show that for $\e={1\042}$ at most one of the contributions
to $\sum \ch_i=3$ can lie in the excluded open intervals
($M_{1\042}$ has 502 elements). Thus a computer
program can produce all $830$ solutions where members of $M_\e$ add up to
$3$ or to $3$ minus a possible value of $\ch$ which lies in the excluded
range. Most of these combinations of $\ch$'s can originate from several
(up to 90)
different potentials of tadpole and Fermat type. The complete number of
inequivalent models of this type
(i.e. $a,e^{(T)}\ge3$, $f^{(T)}\ge2$ and $e^{(L)}\ge f^{(L)}\ge2$) turns
out to be $7579$.
$d$ in (\ref{lgp}) is a multiple of all $a$'s, $({ef-1\0\gcd(e-1,f-1)}
)^{(L)}$'s, and $({ef\0\gcd(e-1,f)})^{(T)}$'s.

All these models are distinct from the orbifold point of view.
They only correspond to 3112 different combinations of weights, however,
which in turn give rise to some 1200 different Hodge pairs.
Considering a specific set of weights, different points in
the configuration space  (\ref{aff})
can have different symmetries, leading to different possibilities
for orbifoldizing. Even taking this into account, our analysis given here
is complete for all potentials in which at most two fields are coupled:
One can show that the potentials (\ref{L},\ref{T}) represent the
points of maximal symmetry in the respective moduli spaces.

\section{Abelian symmetries of LG potentials}

In this section we first determine pure  phase symmetries in the canonical
basis (\ref{chain}-\ref{iiiv}).
Then we discuss the diagonalization of states for
abelian combinations of phase symmetries and cyclic permutations and work
out the formulae for the dimensions and eigenvalues of the eigenspaces for
general group actions. The final ingredients for the calculation of the
chiral ring of the orbifold are the quantum numbers of the vacuum and the
LG description of the twisted sectors.

We want to construct all abelian symmetries which respect quasihomogeneity.
These consist of combinations of cyclic permutations and phase transformations.
We need not consider permutations within a coupled
sector, because due to the resulting restrictions on the
phases we would not get any new models.

Let us first discuss all possible phase groups for
the models (I)--(IV)
of the previous section, generated by transformations $\cp:\Phi_i\to \r_i
\Phi_i$ with $\ln\r_i=2\p i \ph_i$.

In the first two cases we have $\r_{i+1}=\r_i^{-a_i}$ for $i\ge 1$ with
\beq \ph_1^{(I)}={1\0\prod a_i},\qqqd\ph_1^{(II)}={1\0\prod a_i-(-1)^n}\eeq
for type (I) and type (II), respectively.

Type three is more complicated. In general there will be two generators,
because there is no particular $\r_i$ which determines all others.
Invariance of $W$ implies $\r_2=\r_1^{-a}=\r_3^{-c}$ and $\r_2^b=\r_1^p
\r_3^q=1$, from which we conclude $\ph_1={m\0ab},\,\ph_3={n\0bc}$
with $n=m+bj$ and $m$ and $j$ chosen in such a way that
$l=p\ph_1+q\ph_3\in\ZZ$.
The minimal value of $m={ab(lc-jq)\0pc+aq}$ is given by
\beq m_0={ab(c\cap q)\0 abq\cap (pc+aq)}=
     {b-1\0 (b-1)\cap{c\0 q\cap c}},                                    \eeq
where ``$\cap$'' means ``greatest common divisor'' (The equality
of these two expressions can be seen with the help of $eq.$ (\ref{pq}).
We define our first generator $\cp_1$ of the maximal phase symmetry
as the group element determined by $m_0$ and $n_0=m_0+jb$
with any $j$ fulfilling ${q\0c}j+{m_0\0b-1}\in\ZZ$. Taking any group
element, repeated application of $\cp_1$ will produce an element with
$m=0$. The remaining freedom can be described by $\cp_2:\,\r_1=\r_2=1,\,
\ph_3={1\0 c\cap q}$.

Type (IV) is similar to the first two cases: There is only one generator $\cp$
determined by $\r_2=\r_1^{-a}$ and $\r_3=\r_2^{-b}$ with
\beq \ph_1={1\0(a(bc-1))\cap(p+abq)}.                                    \eeq

We now consider additional cyclic symmetries. If we have $n$ copies of the
same Fermat type model in
$W=\sum_{i=1}^n \Phi_i^a+\ldots$, then the maximal
abelian group which mixes all $n$ fields can have 2 generators.
By a linear change of variables these act as
\bea \cc:\Phi_n\to\s \Phi_1,&\;\Phi_i\to\Phi_{i+1}\qd i<n,\qd&\ln \s=2\p
     i{s\0a}\\
     \cp:\Phi_i\to\r \Phi_i,&                           &\ln \r=2\p i{r\0a}\eea
with
\beq \ln\,\det\cc=2\p\i\left({n-1\0 2}+{s\0a}\right),
     \qquad \ln\,\det\cp=2\p\i{nr\0a}.\eeq
Upon diagonalization of $\cc$ we find that its eigenvalues are
equidistant on the unit circle:
\beq\cc\ti \Phi_j=\r_j\ti \Phi_j,\qqd \ph_j={s\0na}+{j\0n},\qd 1\le j\le n.\eeq
It would be complicated to work in the diagonal basis
$\ti \Phi_j={1\0n}\sum_{i=1}^n\r_j^{-i}\Phi_i$, but fortunately
it is not difficult to calculate the dimensions of the eigenspaces, which is
all we need: The number of states for a given degree of homogeneity $h$ in
$\Phi_i$ is
\beq\ca_a(n,h)=\sum_{j=0}^{h\0a-1}(-1)^j{n\bi j}{h+n-1-j(a-1)\bi n-1}.    \eeq
$h$ is an integer between 0 and $n(a-2)$.
For all states $\Phi_1^{\l_1}\ldots \Phi_n^{\l_n}\;$ $\cc^n$ is diagonal. The
states in the orbit of $\cc$ in general yield $n$ diagonalized states
with eigenvalues
\beq{\rm exp}\,2\p\i\left({hs\0na}+{j\0n'}\right),\eeq
$n'=n$.
If the $\l_i$'s have a cyclic symmetry $\l_i=\l_{i+n'}$, then
$\cc^{n'}$ is diagonal for a divisor $n'$ of $n$, with the above
formula for the eigenvalues still being valid. We thus need to calculate
the number of states with definite $h=\sum_{i=1}^n\l_i=g\sum_{i=1}^{n'}\l_i$
and a cyclic symmetry of order $g=n/n'$ in the exponents $\l_i$.
Including a factor $1/n'$ for the number of eigenspaces
into which these states decompose, this number $\tilde\ca({n\0n'},{h\0n'})$
can be calculated recursively by
\beq n\ti\ca_a(n,h)=\ca_a(n,h)-\sum_{1<m|\gcd(n,h)}{n\0m}
                              \ti\ca_a\left({n\0m},{h\0m}\right), \eeq
where $m$ runs over the common divisors of $n$ and $h$.
For a definite $h$ the dimension of an eigenspace with phase
$\th={hs\0na}+{j\0n}$  of the respective eigenvalue of $\cc$ is
\beq\hat\ca_a(n,h,g)=\sum_{\g|g}\ti\ca_a\left({n\0\g},{h\0\g}\right),    \eeq
where $\g$ runs over all divisors (including 1) of $g=\gcd(j,n,h)$ and
$j=n\th-{hs\0a}$ has to be integer.
The phase of the eigenvalue of $\cp$ is, of course, $hr\0a$ modulo 1.

Producing an orbifold by modding out with respect to the symmetries considered
above, we have to pay special attention to twisted states. In a particular
twisted sector, where the string closes (on the considered fields) up to a
group
transformation $\cc^I\cp^J$, only invariant fields contribute to the ground
states in the Ramond sector, and thus to the chiral rings in the Neveu Schwarz
sector. Let $t=\gcd(I,n)$. Then there exist exactly $t$ invariant fields if
the equation $I({s\0an}+{j\0n})+J{r\0a}\aus\ZZ$ has at least one integer
solution $j$, which is the case if and only if
\beq {Is+Jrn\0ta}\aus\ZZ.                                            \lleq{Xtw}
The chiral states can be calculated from an ``effective'' Landau Ginzburg
model with $n\eff=t$ fields and with
\beq s\eff=t(s+j_0a)/n \qd\mod\qd a,                                       \eeq
where $(I(s+j_0a)+Jnr)/(na)$ is integer (we may choose $0\le j_0< {n\0t}$).
Using $\Phi_i=\sum_{j=1}^n\r_j^i\ti \Phi_j$,
\beq W_{\rm eff}=\sum_{i=1}^n(\sum_{k=1}^t\r_{j_0+{kn\0t}}^i
     \ti \Phi_{j_0+{kn\0t}})^a={n\0t}\sum_{i=1}^t \Phi_{i\,\rm eff}^a     \eeq
with
$\Phi_{i\,\rm eff}={n\0t}\sum_{k=0}^{{n\0t}-1}{\rm exp}(-2\p\i \,s_{\rm eff}
      {k\0a})\Phi_{i+tk}$.
According to eq. (\ref{qpm})
the contribution of the non-invariant fields to the charges, i.e. the
left/right charge of the twisted vacuum, is
\beq \D q_\pm=\sum_{\tt i\not\aus\ZZ}(\8\2-q_i\pm(\tt i -\8\2))=
(n-n\eff)(\8\2-q_\Phi)\pm (t-n\eff)\(\left[{Is+Jnr\0ta}\right]-\2\)\eeq
where $n\eff=t=\gcd(n,I)$ if (\ref{Xtw}) is satisfied and  $n\eff=0$ otherwise.

The same analysis can be done for the phase symmetries we found at the
beginning of this chapter, combined with a cyclic symmetry which now permutes
complete coupled sectors. For
type (I) and (II)  models with $n=2$ the results are listed in the remainder
of this section.

For $W^{(L)}=\sum_{i=1}^n(\Phi_i^e\Psi_i+\Psi_i^f\Phi_i)$ the number of
monomials of degree $k$ in $\Phi$ and of degree $l$ in $\Psi$ is
\beq\cL_{ef}(n,k,l)=\ca_{e+1}(n,k)\ca_{f+1}(n,l),                          \eeq
whereas for $W^{(T)}=\sum_{i=1}^n(\Phi_i^e+\Psi_i^f\Phi_i)$ there are
\beq\cT_{ef}(n,k,l)=\sum_{j\le n}
                     {n\choose j}\ca_{e+1}(j,k)\ca_f(j,l-(n-j)(f-1))       \eeq
monomials of degree $k$ in $\Phi$ and of degree $l$ in $\Psi$. As before,
it is useful to define
\bea\cL_{ef}(n,k,l)    &&\eq\sum_{m|\gcd(n,k,l)}
                         {n\0m}\ti\cL_{ef}({n\0m},{k\0m},{l\0m})\\
  \hat\cL_{ef}(n,k,l,g)&&\eq\sum_{\g|g}\ti\cL_{ef}({n\0\g},{k\0\g},{l\0\g})\eea
and the analogous quantities for $\cT$.
$\hat\cL(n,k,l,g)$ and $\hat\cT(n,k,l,g)$ are the dimensions of the eigenspaces
with eigenvalue $\exp(2\p \i\th)$ of $\cc$, where $j=n\th-s{l-fk\0\co}$
has to be integer, $g=\gcd(n,k,l,j)$ and $\co=ef-1$  are $\co=ef$
the orders of the respective maximal phase symmetries.
The phases of the eigenvalues of $\cp$ are $r(l-fk)/\co$ and
the determinants are given by
\beq \ln\,\det\cc=2\p\i{s(1-f)\0\co},\qquad
\ln\,\det\cp=2\p\i{nr(1-f)\0\co}.\eeq
These formulae are valid for both loops and tadpoles.

Now we  consider the twisted sector for a group element $\cc^I\cp^J$.
$t\eq \gcd(n,I)$ pairs of fields $(\Phi,\Psi)$ contribute to the chiral ring in
this sector iff
\beq {sI+nJr\0t\co}\aus\ZZ,                   \label{surv}            \eeq
implying charges
\beq \D q_\pm=(n-n\eff)(1-q_\Phi-q_\Psi)\pm t\(\left[{Is+Jnr\0
             t\co}\right]-\left[{f(Is+Jnr)\0 t\co}\right]\) \label{dpm} \eeq
of the twisted vacuum (in this case $n\eff=t$). If $eq.$ (\ref{surv}) is not
fulfilled, the chiral ring only consists of the twisted vacuum and
(\ref{dpm}) is still valid, but now $n\eff=0$.
The effective LG theory describing the twisted sectors has
\beq s\eff=t(s+j_0\co)/n \mod\co,                                        \eeq
where $j_0$ is determined by $(I(s+j_0\co)+Jnr)/(n\co)\aus\ZZ$.

A peculiarity of the tadpole type is that even if $\Psi$ is not invariant,
$\Phi$
still can be invariant. Thus, if equation (\ref{surv}) is not satisfied, but
\beq {sI+nJr\0et}\aus\ZZ,                                                 \eeq
then the effective LG-theory is of Fermat type with $n\eff=t$, $a\eff=e$,
$r\eff=-r \mod e$ and
\beq s\eff=t(-s+j_0e)/n \mod   e                                          \eeq
with $(j_0Ie-Is-Jnr)/(ne)\aus\ZZ$ and
\beq \D q_\pm=n(1-q_\Phi-q_\Psi)+t(q_\Phi-\8\2)
        \pm t\(\left[{Is+Jnr\0 tef}\right]-\2\).                         \eeq

\section{Actions of Symmetries: General Considerations}
In this section we will discuss some general aspects that are important
for group actions on Landau--Ginzburg theories that have been
orbifolded with respect to the U(1)--symmetry in order to describe
string vacua with $N=1$ spacetime supersymmetry.

An obvious question when considering  orbifolds is whether
there is any a priori insight into
what spectra are possible for the orbifolds of a given model with respect
to a particular set of symmetries.
This question is of particular interest if the goal is to produce
orbifolds with presribed spectra, say models with a small number
of fields where the difference between the number of generations and
antigenerations is three.

Even though it is possible to formulate  constraints on the
orbifold spectrum for
particular types of actions, we know of no constraints that hold
in full generality, or even for arbitrary cyclic actions.
One very simple class  of symmetries  are those without fixed points.
For such actions there are no twisted sectors and hence
there exists a simple formula expressing the Euler number $\chi_{orb}$ of the
orbifold in terms of the Euler characteristic $\chi$ of the covering space
and the order $|G|$ of the group
\beq
\chi_{orb} = \frac{\chi}{|G|}.
\eeq
The vast majority of actions however do have fixed points and hence the
result above does not apply very often.

For orbifolds with respect to cyclic groups of prime order there
exists a generalization of this result. For such group actions it was shown
in \cite{d} that
\beq
\bn^g_{orb} - n^g_{orb} =
\left( |G|+1\right)\left(\bn^g_{inv} - n^g_{inv}\right)
-\left(\bn^g - n^g\right),
\eeq
where $n^g_{orb}$, $n^g_{inv}$, $n^g$ are the numbers of generations
of the orbifold theory, the invariant sector and the original LG theory,
respectively.

Consider then the problem of constructing an orbifold with a prescribed Euler
number $\chi_{orb}$ from a given theory.
Only for fixed point free actions will the order of the group be completely
specified as $|G|=\chi/\chi_{orb}$. It is important to realize that in
general the order of the group by which a theory is orbifolded does
{\it not} determine its spectrum -- the precise form of the action of the
symmetry is important.

Nevertheless we can derive {\it some} constraints on the order of the
action that we are looking for. Even though we don't  know a priori what
the invariant
sector of the orbifold will be we do know that its associated Euler number
must be an integer
\beq
\chi_{inv} = \frac{\chi + \chi_{orbi}}{|G|+1}~~\in \IN.
\eeq
This simple condition does lead to restrictions for the order of the group.
Suppose, e.g., that we wish to check whether the quintic threefold admits
a three--generation orbifold: For the deformation class of the quintic
\beq
\IC_{(1,1,1,1,1)}[5]: ~\chi=-200
\eeq
the order of the discrete group in question must satisfy the constraint
$-206/(|G|+1) \in \ZZ$,  implying $|G|=102$.
Hence there exists no three--generation orbifold of the quintic with respect
to a discrete group with prime order.
A counterexample for nonprime orders is furnished by the following theory
\beq
\IC_{(2,2,2,3,3,3,3)}[9]:~ (\bn^g,n^g,\c)=(8, 35,-54),
\lleq{3gen}
which corresponds to a CY theory embedded in a product of two projective
spaces by two polynomials of bidegree $(0,3)$ and $(3,1)$ \cite{s2}
\footnote{i.e. the Calabi--Yau manifold of this model is embedded in
           an ambient space consisting of a product of two projective spaces
{\footnotesize $\matrix{\IP_2\cr \IP_3\cr} \left[\matrix{3&0\cr 1&3\cr}
\right]$}.}.
Suppose we are searching for three--generation orbifolds of this space
with $\chi_{orb}=\pm 6$ . If $\chi=-6$ the constraint is not very restrictive
and allows a number of possible groups $|G|\in \{2,3,5,11,19,29\}$.
Even though it is not known whether any of these groups lead to a
three--generation model it {\it is}  known that at a particular
point in the configuration space of (\ref{3gen}) described by the
superpotential
\beq  W=\sum_{i=1}^3 (\Phi_i^3+\Phi_i \Psi_i^3)+\Phi_4^3   \eeq
a symmetry of order nine exists that leads to a three--generation model
\cite{s2}.

Our interest however is not restricted to models with particular spectra
for reasons explicated in the introduction. Hence we wish to implement
general types of actions regardless of their fixed point structure and
order. A general analysis of symmetries for an arbitrary Landau--Ginzburg
potential is beyond the scope of this paper; instead we restrict our
attention to the types of potentials that we have constructed explicitly.
Before we discuss these types we should remark upon a number of aspects
concerning actions on string vacua defined by LG--theories.

It is important to note that depending on the weights (or charges) of the
original LG theory it can and does happen that actions that take rather
different forms when considered as actions
on the LG theory actually are isomorphic when viewed as action of the
string vacuum proper because of the U(1) projection. It is easiest to
explain this with an example.
Consider the superpotential
\beq
W=\Phi_1^{18}+\Phi_2^{18}+\Phi_3^3+\Phi_4^3+\Phi_4\Phi_5^3
\eeq
which belongs to the configuration
$\IC_{(1,1,6,6,4)}[18]_{-204}^9$ (here the superscript denotes the
number of antigenerations and the subscript denotes the Euler number of the
configuration). At this
particular point in moduli space we can, e.g., consider the orbifolds with
respect to the actions
\bea
\ZZ_3&:&\left[\matrix{0&0&1&0&2\cr}\right],~~~(13, 79, -132) \nn \\
\ZZ_3&:&\left[\matrix{1&1&1&0&0\cr}\right],~~~(13, 79, -132) \nn \\
\ZZ_3&:&\left[\matrix{1&0&1&0&1\cr}\right],~~~(14, 44, -60) ,\label{z3}
\eea
where the notation $\ZZ_a: [p_1~\ldots~p_n]$ indicates that the
fields $\Phi_i$ transform with phases $(2\pi i p_i/a)$ under the
generator of the $\ZZ_a$ symmetry.
It is clear from the last action in (\ref{z3}) that the order of a group
is, in general, not suffient to determine the resulting orbifold spectrum
but that the specific form of the way the symmetry acts is essential.

Since the first two actions lead to the same spectrum we are led to ask
whether the two resulting orbifolds are equivalent.
Theories with the
same number of light fields need, of course, not be equivalent and to
show whether they are is, in general a rather involved analysis,
entailing the transformation behaviour of the fields and the computation
of the Yukawa couplings.

In the case at hand it is, however, very easy to check this question.
The first two actions only differ by the $6^{\rm th}$ power of the
canonical $\ZZ_{18}$ which is given by $\ZZ_3:\,[1~1~0~0~1]$.
Since the orbifolding with respect to this group is always present
in the construction of an LG vacuum the fist two orbifolds in
$eq.$~(\ref{z3}) are trivially equivalent.

Another important point is the role of trivial factors in the LG theories.
Given a superpotential $W_0$ with the correct central charge to define a
Heterotic String vacuum we always have the freedom to add trivial factors to it
\beq
W=W_0 + \sum_i \Phi_i^2,
\eeq
since neither the central charge nor the chiral ring are changed by this
operation.
 As we restrict our attention to symmetries with unit determinant,
we gain, however, the possibility to cancel a negative sign of the determinant
by giving some $\Phi_i$ a nontrivial transformation property under a
$\ZZ_{2n}$. Adding a trivial factor hence changes the symmetry properties
of the LG--potential with regards to this class of symmetries.
\footnote{In LG theories the determinant restriction is necessary for modular
   invariance and can be avoided by introducing discrete torsion \cite{iv}.}
If we wish to relate the vacuum described by the potential to a
Calabi--Yau manifold, consideration of trivial factors becomes essential
\cite{ls1}.
Consider e.g. the LG--potential
\beq
W_0=\Phi_1^{12}+\Phi_2^{12}+\Phi_3^6+\Phi_4^6
\eeq
which has $c=9$ and charges
$(\frac{1}{12},\frac{1}{12},\frac{1}{6},\frac{1}{6})$ and hence is a member
of the configuration  $\IC_{(1,1,2,2)}[12]$. Only after adding the necessary
trivial factor
this theory can be orbifolded with an action defined by
$\ZZ_2:[~1~0~0~0~1~]$ acting on the Fermat polynomial in
$\IC_{(1,1,2,2,6)}[12]$; this action leads to the orbifold spectrum
$(4,94,-180)$ and is not equivalent to any symmetry that
acts only on the first four variables  with determinant 1. Neglecting the
addition of the quadratic term to the LG potential $W_0$
would have meant missing the above spectrum as one of the possible
orbifold results.

Finally it should be noted that obviously we have to make {\it some}
choice about which points in moduli space we wish to consider. Different
members of a moduli space have, in general, drastically different
symmetry properties. An example is the well known quintic theory
which we already mentioned. The most symmetric
point in the 101 dimensional space of complex deformations of the quintic is
described by the Fermat polynomial
\beq W= \sum_i \Phi_i^5,                                              \eeq
which has a discrete symmetry group of order $5!\cdot 5^4$. Any deformation
breaks most of these symmetries. It turns out that for the quintic
the different points in moduli space that we have considered
do not lead to actions which provide
additional spectra. In other examples it does happen that the consideration
of additional points leads to new result. The configuration
$\IC_{(15,3,2,20,20)}[60]_{-48}^{31}$ e.g.
admits Fermat type polynomials as well as tadpole type polynomials, and
even though none of the Fermat actions we have implemented leads to a
three--generation model the following tadpole--Fermat type potential
\beq
W=\Phi_1^4+\Phi_2^{20}+\Phi_3^{30}+\Phi_4^3+\Phi_4\Phi_5^2
\eeq
leads to a model with spectrum
$(36, 39, -6)$ when orbifolded with the symmetry  $\ZZ_2~:[~1~0~0~0~1]$.

\noindent
\section{Phase Actions: Implementation and Results}

Consider then a potential $W$ with $n$ order parameters
normalized such that the degree $d$ takes the lowest value such that all order
parameters have integer weight.
In the following we discuss potentials of the type
\beq
W= \sum_i \Phi_i^{a_i} + \sum_j \left(\Phi_j^{e_j} + \Phi_j \Psi_j^{f_j}\right)
   + \sum_k \left(\Phi_k^{e_k}\Psi_k + \Phi_k \Psi_k^{f_k} \right)
\eeq
which consist of Fermat parts, tadpole parts and loop parts.

\noindent
{\bf FERMAT POTENTIALS}:
Clearly the potential $W=\sum_{i=1}^n \Phi_i^{a_i}$ is
invariant under $\prod_i \ZZ_{a_i}$, i.e. the phases of the individual fields,
acting like
\beq
\Phi_i \longrightarrow e^{2\pi i \frac{m_i}{a_i}} \Phi_i.
\eeq
For some divisor $a$ of ${\rm lcm}(a_1,\dots, a_n)$ and ${m_i\0a_i}={p_i\0a}$
we denote such an action by
\beq
\ZZ_a:~\left[\matrix{p_1&p_2 &\cdots &p_n\cr}\right],~~~0\leq~ p_i \leq~ a-1.
\eeq
and require that $a$ divides $\sum p_i$ in order to have determinant 1.

We have implemented such symmetries in the form
\beq
\ZZ_a:~\left[\matrix{(a-\sum_l i_l)&i_1&\cdots&i_p
             &(a-\sum_m j_m)&j_1&\cdots&j_q&\cdots} \right]     \eeq
with the obvious divisibility conditions. For small $p$ and $q$
these symmetries can act on a large number of spaces and therefore lead to
many different orbifolds, but as $p,q$ get larger
the number of resulting orbifolds decreases rapidly.
We have stopped implementation of more complicated actions when the number of
results for the different orbifold Hodge pairs was of the order of a few tens.
As already mentioned above, the precise form of the action is very important
when considering symmetries with fixed points since the order itself is
not sufficient to determine the orbifold spectrum.

More complicated symmetries can be constructed via multiple actions
by multiplying single actions of the type described above
\beq
\prod_c \ZZ_{a_c} :~~
\left[\matrix{(a_c-\sum_l i_{c,l})&i_{c,1}&\cdots&i_{c,p}
             &(a_c-\sum_m j_{c,m})&j_{c,1}&\cdots&j_{c,q}&\cdots} \right].
\eeq
We have considered (an incomplete set of) actions of this type with up to
six twists (i.e. six $\ZZ_a$ factors). Again the precise form of the action
is rather important.

\vskip .2truein

\noindent
{\bf TADPOLE AND LOOP POLYNOMIALS}:~~
The action of the generator of the maximal phase symmetry within a tadpole
or loop sector is
\beq \ZZ_{\co}: \left[\matrix{-f&1}\right],                             \eeq
where $\co = ef$ or $ef-1$, respectively. If we want unit determinant within
one sector, we must take our generator to the $n^{\rm th}$ power with some
$n$ fulfilling $n(f-1)/\co\in\ZZ$. With $\o =gcd(f-1,\co)$ the action
of the resulting subgroup can be chosen to be
\beq
\ZZ_{\o}:~~ \left[\matrix{(\o-1)&1}\right].
\eeq

Other types of actions that we have considered for superpotentials consisting
of Fermat parts and tadpole/loop parts involve phases acting both on the
tadpole/loop part as well as on a number of Fermat monomials.
As was the case with pure Fermat polynomials we have also implemented
multiple actions of the type considered above.

We have implemented some forty different actions of the types
described in the previous paragraphs. These symmetries lead to a large
number of orbifolds not all of which are distinct however for reasons
explained in the previous section.
Our computations have concentrated on the number of generations and
anti--generations of these models and we have found some 2000 distinct
Hodge pairs
\footnote{This number is very close to the number of spectra found in
           \cite{cls}
           for the number of distinct Hodge pairs in a large class of
           LG--theories equivalent to CY manifolds embedded in weighted
           $\IP_4$}.
In Fig. 1 we have plotted the difference of the number of generations and
antigenerations versus their sum for all the Hodge pairs.

It is obvious from this plot that there is a large overlap between the
results of \cite{cls} and the orbifolds constructed here. This might indicate
that the relation established in \cite{ls2} between orbifolds of
Landau--Ginzburg
theories and other Landau--Ginzburg theories is a general phenomenon and
not restricted to the particular classes of actions which were analysed in
\cite{ls2}.

Models with a low number of fields are clearly
of particular interest. There are two aspects to this question,
as mentioned in the introduction -- low numbers for the {\it difference} of
generations and anti--generations (more precisely one wants the number 3 here)
and low values for the {\it total} number of generations and anti--generations.
As far as the latter are concerned the following `low--points' are the
`highlights' among the results for phase symmetry orbifolds.

The lowest models have $\chi=0$, more precisely the spectra (9,9,0)
and (11,11,0). These spectra appear many times in different
orbifolds of Fermat type; an example for the first one being
\beq
\IC_{(1,\ldots,1)}[9]/\ZZ_3^2: \left[\matrix{1&1&1&0&0&0&0&0&0\cr
                           0&0&0&1&1&1&0&0&0\cr}\right]        \eeq
or, even simpler,
\beq
\IC_{(4,4,4,4,4,4,3,3)}[12]/\ZZ_3: [1~1~1~0~0~0~0~0].
\eeq
The second one can be constructed e.g. as
\beq
\IC_{(4,3,3,3,3,2)}[12]/\ZZ_4^2: \left[\matrix{0&1&1&2&0&0\cr
                                               0&0&2&1&1&0\cr}\right].
\eeq

Other examples with a total of 22 generations and anti--generations are
the following orbifolds of the Fermat quintic:
\beq
\ZZ_5: \left[\matrix{0&1&2&3&4}\right],~~\qqd (1,21,-40)
\eeq
and
\beq
\ZZ_5^2: \left[\matrix{3&1&1&0&0\cr
              0&3&1&1&0\cr}\right],~~\qqd (21,1,40)
\eeq

\noindent
Of particular interest, of course, are three--generation models.
In the list of 3112 models there are, aside from the known \cite{cls}
three--generation models embedded in weighted $\IP_4$, no new three
generation models. For completeness we list these models in Table 1.

\noindent
{\it Table 1.} Three--generation LG models.

\begin{center}
\begin{small}
\begin{tabular}{||l l l||}
\hline
\hline
Configuration &  &Potential \tabroom \\
\hline
\hline
$\IC_{(21,21,14,3,4)}[63]_{-6}^{32}$ &\qqd
&$\Phi_1^3+\Phi_2^3+\Phi_2\Phi_3^3+\Phi_4^{21}+\Phi_4\Phi_5^{15}$
							       \tabroom \\
\hline
\hline
$\IC_{(15,5,8,3,14)}[45]_{-6}^{20}$ &\qqd
&$\Phi_1^3+\Phi_2^9+\Phi_2\Phi_3^5+\Phi_4^{15}+\Phi_4\Phi_5^3$\tabroom \\
\hline
\hline
$\IC_{(17,6,9,17,2)}[51]_6^{34}$ &\qqd
&$\Phi_1^3+\Phi_2^7\Phi_3+\Phi_2\Phi_3^5+\Phi_4^3+\Phi_4\Phi_5^{17}$
								   \tabroom \\
\hline
\hline
\end{tabular}
\end{small}
\end{center}

Via orbifolding we find a number of such models which all however have
a fairly large number of generations and antigenerations. We list those
in Table 2.

\noindent
{\it Table 2}. Three--generation orbifold models: we do not list models
separately which are equivalent up to the U(1) projection.

\begin{small}
\begin{center}
\begin{tabular}{||l|l l l r||}
\hline
\hline
$\#$ &Configuration  &Potential                             &Action
&Spectrum \tabroom \\
\hline
\hline
1 &$\IC_{(9,2,5,9,2)}[27]_{-66}^{16}$
&$\Phi_1^3+\Phi_2^{11}\Phi_3+\Phi_2\Phi_3^5+\Phi_4^3+\Phi_4\Phi_5^9$
&$\ZZ_3~:[1~0~0~0~2]$                  &$(18, 21, -6)$ \tabroom \\ [2ex]
2 &$\IC_{(17,6,9,3,16)}[51]_{-102}^{15}$
&$\Phi_1^3+\Phi_2^7\Phi_3+\Phi_2\Phi_3^5+\Phi_4^{17}+\Phi_4\Phi_5^3$
&$\ZZ_2~:[0~1~1~0~0]$                  &$(31, 34, -6)$ \\ [2ex]
3 &$\IC_{(9,2,5,3,8)}[27]_{-54}^{10}$
&$\Phi_1^3+\Phi_2^{11}\Phi_3+\Phi_2\Phi_3^5+\Phi_4^9+\Phi_4\Phi_5^3$
&$\ZZ_2~:[0~1~1~0~2]$                  &$(21, 18, 6)$ \\ [2ex]
4 &$\IC_{(15,15,2,9,4)}[45]_{-30}^{23}$
& $\Phi_1^3+\Phi_2^3+\Phi_2\Phi_3^{15}+\Phi_4^5+\Phi_4\Phi_5^9$
&$\ZZ_3~:[1~0~2~0~0]$                  &$(23, 20, 6)$\\ [2ex]
5 &$\IC_{(15,15,10,3,2)}[45]_{-54}^{22}$
&$\Phi_1^3+\Phi_2^3+\Phi_2\Phi_3^3+\Phi_4^{15}+\Phi_4\Phi_5^{21}$
&$\ZZ_3~:[1~0~2~0~0]$                  &$(35, 32, 6)$ \\ [2ex]
\hline
\hline
\end{tabular}
\end{center}
\end{small}

\noindent
By using the relation established in \cite{ls2} between LG/CY--theories
via fractional transformations it can be shown that the orbifold $\#1$ in
Table 2,
\beq
\IC_{(2,5,9,2,9)}[27]_{-66}^{16}/\ZZ_3:[~0~0~0~2~1],
\eeq
for which the covering model is described by the polynomial
\beq
W= \Phi_1^{11}\Phi_2+\Phi_1\Phi_2^5+\Phi_3^3+\Phi_3\Phi_4^9+\Phi_5^3,
\eeq
is isomorphic to the orbifold
\beq
\IC_{(2,5,9,3,8)}[27]_{-54}^{10}/\ZZ_2:[~0~0~0~1~1]
\eeq
where the covering theory is described by the polynomial
\beq
W= \Phi_1^{11}\Phi_2+\Phi_1\Phi_2^5+\Phi_3^3+\Phi_3\Phi_4^6
                                                     +\Phi_4\Phi_5^3.
\eeq
The latter is a theory involving a subtheory with couplings among three
scaling fields and hence goes beyond the types of potentials we have
implemented.
This example indicates that more complicated examples than the ones
investigated here are likely to yield more (perhaps more realistic)
three generation models.

The covering spaces of all the three generation models are described by
either tadpole or loop type polynomials, and with our actions none of the
Fermat type polynomials leads to a three generation model.
It should be noted that these orbifolds exist only at particular points in
moduli space. In some cases the tadpole polynomial defining the covering
space configuration admits a Fermat representation, but it turns out that
this is not the point in moduli space that leads to a three generation model.

\noindent
\section{Cyclic Permutations}

Consider a cyclic permutation of order $r=2n$ generated by
\beq
(\Phi_1,\Phi_2, \dots ,\Phi_r)\mapsto
(\Phi_2, \dots ,\Phi_r,\Phi_1).
\eeq
Such an action is not allowed since its determinant is $-1$.
The direct sum of an even number of such permutations is, of course,
a good symmetry.

Since the total number of fields is at most nine, there is only a small
number of possible pure cyclic permutations.
The implementation of all these cyclic permutations leads to several
hundred orbifolds which however lead only to some 100 different Hodge pairs
which we plot in Figure 2.


It should be noted that the model with the smallest number of particles
among all our orbifolds is in this set; it has the  spectrum $(0,12,-24)$ and
comes from the cyclic permutation orbifold
\beq
\IC_{(1,\ldots,1)}[9]/ \ZZ_{9,cyclic} : (0,12,-24).
\eeq
The model with the next smallest number of fields is the $\ZZ_3$
permutation orbifold of the theory described by the potential
\beq W=\Phi^3+\sum_{i=1}^3(\Phi_i^3+\Phi_i\Psi^3_i)                 \eeq
in the configuration $\IC_{(3,3,3,3,2,2,2)}[9]^8_{-54}$ mentioned already.
The cyclic
symmetry permutes the pairs of fields $(\Phi_i, \Psi_i)$
for $i=1,2,3$ and leads to a theory with the spectrum $(4,13,-18)$.
Another model with a comparatively low number
of fields is the Fermat orbifold
\beq
\IC_{(1,\ldots,1)}[9]/ \ZZ_{7,cyclic} : (6,18,-24).
\eeq

\section{Conclusions}

Using the methods we have worked out for the computation of the spectra
of a large class of abelian orbifolds, we have performed a computer search
for all potentials consisting of a superposition of polynomials in
which at most two fields are coupled.

For pure phase symmetries we observe a large overlap with canonically
orbifolded LG vacua, but we also find additional models, in particular
with a small total number of fields.
New three generation orbifolds are found  only for tadpole and loop type
theories.
There are also some models with a relatively low number of fields.

It is clear from our results that the most promising avenue to produce
phenomenologically interesting models is to consider orbifolds with
mixed actions of phase symmetries and permutations since these models
likely lead to spectra that will populate the lower part of the plot.
Generalizing the implementation of cyclic permutations to arbitrary
permutations is also of interest for the breaking of the $E_6$ gauge symmetry
present in this class of theories. Nonabelian symmetries allow to recover
the gauge symmetries of the Standard Model.

As mentioned previously our construction of potentials and implementation of
symmetries is not complete. A very rough measure of the completeness of
our implementation can be gained by considering the orbifold `descendents'
of particular models. Consider e.g. the Fermat potential
\beq
W= \sum_{i=1}^3 \Phi_i^{10}+\Phi_4^5+\Phi_5^2
\eeq
in $\IC_{(1,1,1,2,5)}[10]_{-288}^1$. For this theory our code produced
a completely mirror symmetric space of orbifold `descendents' which we list
in Table 3
\footnote{the pair $(7, 67, -120)$ and $(67, 7, 120)$ was missed in ref.
	  \cite{gp}.}.       \\
\noindent \vbox{
{\it Table 3.} Orbifolds of the Fermat theory in
$\IC_{(1,1,1,2,5)}[10]_{-288}^1$.
\nobreak
\def\tabroom{\hbox to0pt{$\matrix{\cr\cr}$\hss}}
\def\troom{\hbox to0pt{$\matrix{\cr\cr\cr}$\hss}}
\begin{center}
\begin{tabular}{||c c c||}
\hline
\hline
Group   &Action  &Spectrum  \tabroom \\
\hline
\hline
$\ZZ_2$ &$ \left[\matrix{1 &1 &0 &0 &0\cr}\right]$ & \qd$(3, 99, -192)$\qd
                                          \tabroom \\
$\ZZ_{10}^2$ &$ \left[\matrix{8 &1 &1 &0 &0\cr
                              0 &9 &1 &0 &0\cr}\right]$  &$(99, 3, 192)$
 \troom \\
\hline
$\ZZ_2^2$ &$ \left[\matrix{1 &1 &0 &0 &0\cr
			   0 &1 &1 &0 &0\cr}\right]$  &$(7, 67, -120)$
 \troom \\
$\ZZ_5^2$ &$ \left[\matrix{4 &1 &0 &0 &0\cr
			   0 &0 &4 &1 &0\cr}\right]$  &$(67, 7, 120)$
 \troom \\
\hline
$\ZZ_5$ &$\left[\matrix{0 &0 &4 &1 &0\cr}\right]$         &$(11, 47, -72)$
 \tabroom \\  \qd
$\ZZ_{10}\times \ZZ_2$\qd &$ \left[\matrix{9&1&0&0&0\cr
                                        0&1&1&0&0\cr}\right]$ &$(47, 11, 72)$
 \troom \\
\hline
$\ZZ_5$ &$ \left[\matrix{4 &1 &0 &0 &0\cr}\right]$       &$(13, 37, -48)$
 \tabroom \\
$\ZZ_{10}\times \ZZ_2$ &$\left[\matrix{8&1&1&0&0\cr
                                       0&0&1&0&1\cr}\right]$  &$(37, 13, 48)$
 \troom \\
\hline
$\ZZ_{10}$ &$ \left[\matrix{9 &1 &0 &0 &0\cr}\right]$       &$(15, 39, -48)$
 \tabroom \\
$\ZZ_{10}$ &$ \left[\matrix{8 &1 &1 &0 &0\cr}\right]$       &$(39, 15, 48)$
 \tabroom \\
\hline
$\ZZ_{10}$ &$ \left[\matrix{7 &2 &1 &0 &0\cr}\right]$       &$(17, 29, -24)$
 \tabroom \\
$\ZZ_{10}$ &$ \left[\matrix{5 &4 &1 &0 &0\cr}\right]$       &$(29, 17, 24)$
 \tabroom \\
\hline
$\ZZ_{10}^2$ &$ \left[\matrix{9 &1 &0 &0 &0\cr
                              0 &9 &1 &0 &0\cr}\right]$     &$(145, 1, 288)$
 \troom \\
\hline
\hline
\end{tabular}
\end{center}    }
There are, however, in our list a fair number of LG--potentials where we
have not yet found a completely mirror symmetric `descendant' space of
orbifolds.

\vfill\eject

\ve




\parskip=2pt    
\def\del#1{#1}  

\def\ve{\vfil\eject} \let\bsk=\bigskip
\def\plot#1#2{\vskip\parskip
		  \vbox{\hrule width\hsize
			\hbox{\kern-0.2pt\vrule height#1
			      \vbox{\hfill}\kern-0.6pt
			      \vrule}\hrule width\hsize}
    \setbox0=\hbox{#2} \dimen0=\wd0 \divide\dimen0 by 2
    \setbox0=\hbox{\kern-\dimen0 #2}
    \dimen3=#1}

\def\hmark{\kern-0.2pt\lower10pt\hbox{\vrule height 5pt}}
\def\leftscalemark{\vbox{\hrule width5pt}}
\def\rightscalemark{\kern-5pt\vbox{\hrule width5pt}}

\def\Place#1#2#3{
    \count10=#1 \advance\count10 by 960
    \dimen1=\hsize \divide\dimen1 by 1920 \multiply\dimen1 by \count10
    \dimen2=\dimen3 \divide\dimen2 by 550 \multiply\dimen2 by #2
    \vbox to 0pt{\kern-\parskip\kern-20truept\kern-\dimen2
    \hbox{\kern\dimen1#3}\vss}\nointerlineskip}

\def\.#1#2{\Place{#1}{#2}{\copy0}}

\def\frame{\Place{-960}{50}{\leftscalemark~~50}
\Place{-960}{100}{\leftscalemark~~100}
\Place{-960}{150}{\leftscalemark~~150}
\Place{-960}{200}{\leftscalemark~~200}
\Place{-960}{250}{\leftscalemark~~250}
\Place{-960}{300}{\leftscalemark~~300}
\Place{-960}{350}{\leftscalemark~~350}
\Place{-960}{400}{\leftscalemark~~400}
\Place{-960}{450}{\leftscalemark~~450}
\Place{-960}{500}{\leftscalemark~~500}
\Place{960}{50}{\rightscalemark\vphantom{0}}
\Place{960}{100}{\rightscalemark\vphantom{0}}
\Place{960}{150}{\rightscalemark\vphantom{0}}
\Place{960}{200}{\rightscalemark\vphantom{0}}
\Place{960}{250}{\rightscalemark\vphantom{0}}
\Place{960}{300}{\rightscalemark\vphantom{0}}
\Place{960}{350}{\rightscalemark\vphantom{0}}
\Place{960}{400}{\rightscalemark\vphantom{0}}
\Place{960}{450}{\rightscalemark\vphantom{0}}
\Place{960}{500}{\rightscalemark\vphantom{0}}
\Place{-960}{0}{\hmark\lower18pt\hbox{-960}}
\Place{-720}{0}{\hmark\lower18pt\hbox{-720}}
\Place{-480}{0}{\hmark\lower18pt\hbox{-480}}
\Place{-240}{0}{\hmark\lower18pt\hbox{-240}}
\Place{0}{0}{\hmark\lower18pt\hbox{0}}
\Place{240}{0}{\hmark\lower18pt\hbox{240}}
\Place{480}{0}{\hmark\lower18pt\hbox{480}}
\Place{720}{0}{\hmark\lower18pt\hbox{720}}
\Place{960}{0}{\hmark\lower18pt\hbox{960}}
\Place{-720}{550}{\hmark}
\Place{-480}{550}{\hmark}
\Place{-240}{550}{\hmark}
\Place{0}{550}{\hmark}
\Place{240}{550}{\hmark}
\Place{480}{550}{\hmark}
\Place{720}{550}{\hmark}
\Place{960}{550}{\hmark}
}

\begin{center}
\plot{8truein}{\tiny{$\bullet$}}
\nobreak
\frame
\nobreak
\.{  -72}{  36}
\.{ -108}{  54}
\.{ -168}{  84}
\.{ -180}{  90}
\.{  -40}{  22}
\.{  -72}{  38}
\.{  -96}{  50}
\.{ -104}{  54}
\.{ -108}{  56}
\.{ -120}{  62}
\.{ -144}{  74}
\.{ -152}{  78}
\.{ -168}{  86}
\.{ -200}{ 102}
\.{ -204}{ 104}
\.{ -288}{ 146}
\.{ -296}{ 150}
\.{  -72}{  40}
\.{ -108}{  58}
\.{ -112}{  60}
\del{               
\.{ -120}{  64}
\.{ -144}{  76}
\.{ -168}{  88}
\.{ -176}{  92}
\.{ -186}{  97}
\.{ -208}{ 108}
\.{ -216}{ 112}
\.{ -240}{ 124}
\.{ -252}{ 130}
\.{ -260}{ 134}
\.{ -540}{ 274}
\.{  -72}{  42}
\.{  -80}{  46}
\.{  -84}{  48}
\.{  -96}{  54}
\.{ -104}{  58}
\.{ -108}{  60}
\.{ -112}{  62}
\.{ -120}{  66}
\.{ -126}{  69}
\.{ -132}{  72}
\.{ -144}{  78}
\.{ -160}{  86}
\.{ -168}{  90}
\.{ -192}{ 102}
\.{ -200}{ 106}
\.{ -204}{ 108}
\.{ -240}{ 126}
\.{ -256}{ 134}
\.{ -324}{ 168}
\.{ -456}{ 234}
\.{ -480}{ 246}
\.{  -72}{  44}
\.{  -84}{  50}
\.{  -96}{  56}
\.{ -120}{  68}
\.{ -132}{  74}
\.{ -144}{  80}
\.{ -180}{  98}
\.{ -196}{ 106}
\.{ -216}{ 116}
\.{ -280}{ 148}
\.{ -288}{ 152}
\.{ -372}{ 194}
\.{ -408}{ 212}
\.{ -420}{ 218}
\.{  -48}{  34}
\.{  -56}{  38}
\.{  -64}{  42}
\.{  -72}{  46}
\.{  -88}{  54}
\.{  -96}{  58}
\.{ -102}{  61}
\.{ -104}{  62}
\.{ -120}{  70}
\.{ -128}{  74}
\.{ -132}{  76}
\.{ -136}{  78}
\.{ -144}{  82}
\.{ -156}{  88}
\.{ -160}{  90}
\.{ -168}{  94}
\.{ -184}{ 102}
\.{ -192}{ 106}
\.{ -232}{ 126}
\.{ -312}{ 166}
\.{ -336}{ 178}
\.{ -360}{ 190}
\.{ -492}{ 256}
\.{  -24}{  24}
\.{  -48}{  36}
\.{  -60}{  42}
\.{  -72}{  48}
\.{  -84}{  54}
\.{  -96}{  60}
\.{ -100}{  62}
\.{ -108}{  66}
\.{ -120}{  72}
\.{ -132}{  78}
\.{ -164}{  94}
\.{ -168}{  96}
\.{ -192}{ 108}
\.{ -216}{ 120}
\.{ -228}{ 126}
\.{ -252}{ 138}
\.{ -264}{ 144}
\.{ -276}{ 150}
\.{ -292}{ 158}
\.{ -348}{ 186}
\.{ -360}{ 192}
\.{ -384}{ 204}
\.{ -444}{ 234}
\.{  -42}{  35}
\.{  -48}{  38}
\.{  -54}{  41}
\.{  -60}{  44}
\.{  -72}{  50}
\.{  -80}{  54}
\.{  -96}{  62}
\.{ -108}{  68}
\.{ -112}{  70}
\.{ -120}{  74}
\.{ -128}{  78}
\.{ -132}{  80}
\.{ -144}{  86}
\.{ -168}{  98}
\.{ -176}{ 102}
\.{ -192}{ 110}
\.{ -240}{ 134}
\.{ -272}{ 150}
\.{ -288}{ 158}
\.{ -528}{ 278}
\.{  -36}{  34}
\.{  -48}{  40}
\.{  -54}{  43}
\.{  -60}{  46}
\.{  -64}{  48}
\.{  -72}{  52}
\.{  -76}{  54}
\.{  -84}{  58}
\.{  -88}{  60}
\.{ -100}{  66}
\.{ -102}{  67}
\.{ -104}{  68}
\.{ -108}{  70}
\.{ -120}{  76}
\.{ -126}{  79}
\.{ -138}{  85}
\.{ -144}{  88}
\.{ -156}{  94}
\.{ -168}{ 100}
\.{ -180}{ 106}
\.{ -192}{ 112}
\.{ -204}{ 118}
\.{ -216}{ 124}
\.{ -228}{ 130}
\.{ -252}{ 142}
\.{ -264}{ 148}
\.{ -276}{ 154}
\.{ -312}{ 172}
\.{ -372}{ 202}
\.{    0}{  18}
\.{  -24}{  30}
\.{  -36}{  36}
\.{  -48}{  42}
\.{  -60}{  48}
\.{  -64}{  50}
\.{  -72}{  54}
\.{  -80}{  58}
\.{  -90}{  63}
\.{  -96}{  66}
\.{ -108}{  72}
\.{ -120}{  78}
\.{ -136}{  86}
\.{ -140}{  88}
\.{ -144}{  90}
\.{ -148}{  92}
\.{ -150}{  93}
\.{ -160}{  98}
\.{ -168}{ 102}
\.{ -176}{ 106}
\.{ -180}{ 108}
\.{ -184}{ 110}
\.{ -192}{ 114}
\.{ -204}{ 120}
\.{ -216}{ 126}
\.{ -228}{ 132}
\.{ -232}{ 134}
\.{ -240}{ 138}
\.{ -288}{ 162}
\.{ -296}{ 166}
\.{ -300}{ 168}
\.{ -624}{ 330}
\.{  -12}{  26}
\.{  -24}{  32}
\.{  -30}{  35}
\.{  -32}{  36}
\.{  -48}{  44}
\.{  -54}{  47}
\.{  -60}{  50}
\.{  -72}{  56}
\.{  -80}{  60}
\.{  -84}{  62}
\.{  -96}{  68}
\.{ -108}{  74}
\.{ -112}{  76}
\.{ -120}{  80}
\.{ -144}{  92}
\.{ -184}{ 112}
\.{ -192}{ 116}
\.{ -216}{ 128}
\.{ -260}{ 150}
\.{ -336}{ 188}
\.{ -368}{ 204}
\.{ -408}{ 224}
\.{ -420}{ 230}
\.{ -456}{ 248}
\.{ -732}{ 386}
\.{    0}{  22}
\.{  -12}{  28}
\.{  -16}{  30}
\.{  -24}{  34}
\.{  -32}{  38}
\.{  -36}{  40}
\.{  -42}{  43}
\.{  -44}{  44}
\.{  -48}{  46}
\.{  -54}{  49}
\.{  -56}{  50}
\.{  -60}{  52}
\.{  -64}{  54}
\.{  -72}{  58}
\.{  -78}{  61}
\.{  -80}{  62}
\.{  -84}{  64}
\.{  -90}{  67}
\.{  -96}{  70}
\.{ -104}{  74}
\.{ -108}{  76}
\.{ -112}{  78}
\.{ -116}{  80}
\.{ -120}{  82}
\.{ -128}{  86}
\.{ -144}{  94}
\.{ -160}{ 102}
\.{ -168}{ 106}
\.{ -180}{ 112}
\.{ -192}{ 118}
\.{ -208}{ 126}
\.{ -240}{ 142}
\.{ -288}{ 166}
\.{ -312}{ 178}
\.{ -324}{ 184}
\.{ -432}{ 238}
\.{ -480}{ 262}
\.{ -528}{ 286}
\.{ -960}{ 502}
\.{  -24}{  36}
\.{  -36}{  42}
\.{  -40}{  44}
\.{  -48}{  48}
\.{  -72}{  60}
\.{  -84}{  66}
\.{  -88}{  68}
\.{  -96}{  72}
\.{ -108}{  78}
\.{ -120}{  84}
\.{ -126}{  87}
\.{ -144}{  96}
\.{ -168}{ 108}
\.{ -204}{ 126}
\.{ -216}{ 132}
\.{ -228}{ 138}
\.{ -240}{ 144}
\.{ -252}{ 150}
\.{ -304}{ 176}
\.{ -348}{ 198}
\.{ -396}{ 222}
\.{ -612}{ 330}
\.{ -900}{ 474}
\.{    0}{  26}
\.{  -12}{  32}
\.{  -24}{  38}
\.{  -32}{  42}
\.{  -36}{  44}
\.{  -48}{  50}
\.{  -72}{  62}
\.{  -88}{  70}
\.{  -96}{  74}
\.{ -108}{  80}
\.{ -120}{  86}
\.{ -132}{  92}
\.{ -144}{  98}
\.{ -152}{ 102}
\.{ -168}{ 110}
\.{ -176}{ 114}
\.{ -192}{ 122}
\.{ -216}{ 134}
\.{ -228}{ 140}
\.{ -264}{ 158}
\.{ -288}{ 170}
\.{ -336}{ 194}
\.{ -376}{ 214}
\.{ -384}{ 218}
\.{ -432}{ 242}
\.{  -12}{  34}
\.{  -24}{  40}
\.{  -32}{  44}
\.{  -36}{  46}
\.{  -42}{  49}
\.{  -44}{  50}
\.{  -48}{  52}
\.{  -60}{  58}
\.{  -72}{  64}
\.{  -80}{  68}
\.{  -96}{  76}
\.{ -104}{  80}
\.{ -108}{  82}
\.{ -120}{  88}
\.{ -128}{  92}
\.{ -138}{  97}
\.{ -156}{ 106}
\.{ -160}{ 108}
\.{ -168}{ 112}
\.{ -180}{ 118}
\.{ -192}{ 124}
\.{ -204}{ 130}
\.{ -228}{ 142}
\.{ -340}{ 198}
\.{ -408}{ 232}
\.{ -456}{ 256}
\.{ -804}{ 430}
\.{   12}{  24}
\.{    0}{  30}
\.{  -12}{  36}
\.{  -16}{  38}
\.{  -24}{  42}
\.{  -40}{  50}
\.{  -48}{  54}
\.{  -60}{  60}
\.{  -64}{  62}
\.{  -72}{  66}
\.{  -80}{  70}
\.{  -96}{  78}
\.{ -100}{  80}
\.{ -102}{  81}
\.{ -108}{  84}
\.{ -112}{  86}
\.{ -120}{  90}
\.{ -144}{ 102}
\.{ -156}{ 108}
\.{ -160}{ 110}
\.{ -168}{ 114}
\.{ -180}{ 120}
\.{ -224}{ 142}
\.{ -240}{ 150}
\.{ -264}{ 162}
\.{ -300}{ 180}
\.{ -336}{ 198}
\.{ -552}{ 306}
\.{ -744}{ 402}
\.{    0}{  32}
\.{  -12}{  38}
\.{  -18}{  41}
\.{  -24}{  44}
\.{  -36}{  50}
\.{  -48}{  56}
\.{  -60}{  62}
\.{  -66}{  65}
\.{  -72}{  68}
\.{  -80}{  72}
\.{  -96}{  80}
\.{ -102}{  83}
\.{ -108}{  86}
\.{ -120}{  92}
\.{ -144}{ 104}
\.{ -152}{ 108}
\.{ -156}{ 110}
\.{ -168}{ 116}
\.{ -192}{ 128}
\.{ -204}{ 134}
\.{ -228}{ 146}
\.{ -264}{ 164}
\.{ -360}{ 212}
\.{   12}{  28}
\.{    0}{  34}
\.{   -8}{  38}
\.{  -12}{  40}
\.{  -16}{  42}
\.{  -24}{  46}
\.{  -32}{  50}
\.{  -36}{  52}
\.{  -40}{  54}
\.{  -42}{  55}
\.{  -48}{  58}
\.{  -56}{  62}
\.{  -64}{  66}
\.{  -72}{  70}
\.{  -80}{  74}
\.{  -96}{  82}
\.{ -104}{  86}
\.{ -108}{  88}
\.{ -120}{  94}
\.{ -132}{ 100}
\.{ -144}{ 106}
\.{ -152}{ 110}
\.{ -156}{ 112}
\.{ -168}{ 118}
\.{ -200}{ 134}
\.{ -216}{ 142}
\.{ -276}{ 172}
\.{ -312}{ 190}
\.{ -336}{ 202}
\.{ -720}{ 394}
\.{   24}{  24}
\.{    0}{  36}
\.{   -6}{  39}
\.{  -12}{  42}
\.{  -16}{  44}
\.{  -24}{  48}
\.{  -36}{  54}
\.{  -48}{  60}
\.{  -60}{  66}
\.{  -64}{  68}
\.{  -66}{  69}
\.{  -80}{  76}
\.{  -84}{  78}
\.{ -108}{  90}
\.{ -112}{  92}
\.{ -120}{  96}
\.{ -132}{ 102}
\.{ -144}{ 108}
\.{ -156}{ 114}
\.{ -168}{ 120}
\.{ -216}{ 144}
\.{ -252}{ 162}
\.{ -276}{ 174}
\.{ -312}{ 192}
\.{ -408}{ 240}
\.{ -456}{ 264}
\.{ -660}{ 366}
\.{   12}{  32}
\.{    0}{  38}
\.{  -12}{  44}
\.{  -16}{  46}
\.{  -20}{  48}
\.{  -24}{  50}
\.{  -32}{  54}
\.{  -36}{  56}
\.{  -48}{  62}
\.{  -52}{  64}
\.{  -60}{  68}
\.{  -72}{  74}
\.{  -96}{  86}
\.{ -120}{  98}
\.{ -128}{ 102}
\.{ -132}{ 104}
\.{ -144}{ 110}
\.{ -168}{ 122}
\.{ -192}{ 134}
\.{ -240}{ 158}
\.{ -256}{ 166}
\.{ -288}{ 182}
\.{ -336}{ 206}
\.{ -672}{ 374}
\.{   12}{  34}
\.{    0}{  40}
\.{   -6}{  43}
\.{  -12}{  46}
\.{  -24}{  52}
\.{  -36}{  58}
\.{  -44}{  62}
\.{  -48}{  64}
\.{  -56}{  68}
\.{  -60}{  70}
\.{  -72}{  76}
\.{  -78}{  79}
\.{  -84}{  82}
\.{  -96}{  88}
\.{ -108}{  94}
\.{ -112}{  96}
\.{ -120}{ 100}
\.{ -132}{ 106}
\.{ -144}{ 112}
\.{ -160}{ 120}
\.{ -168}{ 124}
\.{ -180}{ 130}
\.{ -264}{ 172}
\.{ -312}{ 196}
\.{ -372}{ 226}
\.{ -564}{ 322}
\.{   40}{  22}
\.{   24}{  30}
\.{   12}{  36}
\.{    8}{  38}
\.{    6}{  39}
\.{    0}{  42}
\.{  -12}{  48}
\.{  -16}{  50}
\.{  -24}{  54}
\.{  -30}{  57}
\.{  -40}{  62}
\.{  -48}{  66}
\.{  -72}{  78}
\.{  -84}{  84}
\.{  -96}{  90}
\.{ -108}{  96}
\.{ -120}{ 102}
\.{ -156}{ 120}
\.{ -184}{ 134}
\.{ -192}{ 138}
\.{ -216}{ 150}
\.{ -232}{ 158}
\.{ -240}{ 162}
\.{ -272}{ 178}
\.{ -288}{ 186}
\.{   24}{  32}
\.{   12}{  38}
\.{    0}{  44}
\.{  -12}{  50}
\.{  -16}{  52}
\.{  -24}{  56}
\.{  -48}{  68}
\.{  -54}{  71}
\.{  -60}{  74}
\.{  -64}{  76}
\.{  -72}{  80}
\.{  -96}{  92}
\.{ -120}{ 104}
\.{ -128}{ 108}
\.{ -168}{ 128}
\.{ -216}{ 152}
\.{ -236}{ 162}
\.{ -288}{ 188}
\.{ -456}{ 272}
\.{ -516}{ 302}
\.{   24}{  34}
\.{   16}{  38}
\.{   12}{  40}
\.{    6}{  43}
\.{    0}{  46}
\.{  -12}{  52}
\.{  -16}{  54}
\.{  -18}{  55}
\.{  -24}{  58}
\.{  -30}{  61}
\.{  -48}{  70}
\.{  -56}{  74}
\.{  -60}{  76}
\.{  -64}{  78}
\.{  -72}{  82}
\.{  -84}{  88}
\.{  -96}{  94}
\.{ -120}{ 106}
\.{ -128}{ 110}
\.{ -144}{ 118}
\.{ -156}{ 124}
\.{ -192}{ 142}
\.{ -240}{ 166}
\.{ -256}{ 174}
\.{ -288}{ 190}
\.{ -480}{ 286}
\.{ -624}{ 358}
\.{   24}{  36}
\.{   12}{  42}
\.{    0}{  48}
\.{   -4}{  50}
\.{   -8}{  52}
\.{  -12}{  54}
\.{  -16}{  56}
\.{  -18}{  57}
\.{  -24}{  60}
\.{  -30}{  63}
\.{  -32}{  64}
\.{  -40}{  68}
\.{  -48}{  72}
\.{  -60}{  78}
\.{  -64}{  80}
\.{  -72}{  84}
\.{  -84}{  90}
\.{  -90}{  93}
\.{ -120}{ 108}
\.{ -128}{ 112}
\.{ -132}{ 114}
\.{ -156}{ 126}
\.{ -180}{ 138}
\.{ -300}{ 198}
\.{ -360}{ 228}
\.{ -564}{ 330}
\.{   30}{  35}
\.{   24}{  38}
\.{   18}{  41}
\.{   16}{  42}
\.{   12}{  44}
\.{    0}{  50}
\.{  -12}{  56}
\.{  -24}{  62}
\.{  -36}{  68}
\.{  -40}{  70}
\.{  -48}{  74}
\.{  -72}{  86}
\.{  -96}{  98}
\.{ -120}{ 110}
\.{ -144}{ 122}
\.{ -168}{ 134}
\.{ -200}{ 150}
\.{ -216}{ 158}
\.{ -384}{ 242}
\.{ -432}{ 266}
\.{   24}{  40}
\.{   16}{  44}
\.{   12}{  46}
\.{    4}{  50}
\.{    0}{  52}
\.{   -4}{  54}
\.{  -12}{  58}
\.{  -24}{  64}
\.{  -36}{  70}
\.{  -48}{  76}
\.{  -60}{  82}
\.{  -72}{  88}
\.{  -84}{  94}
\.{ -112}{ 108}
\.{ -120}{ 112}
\.{ -144}{ 124}
\.{ -252}{ 178}
\.{ -264}{ 184}
\.{ -468}{ 286}
\.{ -588}{ 346}
\.{   36}{  36}
\.{   24}{  42}
\.{   16}{  46}
\.{   12}{  48}
\.{    0}{  54}
\.{  -20}{  64}
\.{  -24}{  66}
\.{  -32}{  70}
\.{  -48}{  78}
\.{  -60}{  84}
\.{  -64}{  86}
\.{  -66}{  87}
\.{  -96}{ 102}
\.{ -108}{ 108}
\.{ -112}{ 110}
\.{ -152}{ 130}
\.{ -156}{ 132}
\.{ -192}{ 150}
\.{ -236}{ 172}
\.{ -240}{ 174}
\.{ -276}{ 192}
\.{   24}{  44}
\.{   12}{  50}
\.{    8}{  52}
\.{    4}{  54}
\.{    0}{  56}
\.{   -8}{  60}
\.{  -12}{  62}
\.{  -16}{  64}
\.{  -24}{  68}
\.{  -32}{  72}
\.{  -36}{  74}
\.{  -48}{  80}
\.{  -72}{  92}
\.{  -76}{  94}
\.{  -80}{  96}
\.{  -84}{  98}
\.{ -120}{ 116}
\.{ -168}{ 140}
\.{ -216}{ 164}
\.{ -228}{ 170}
\.{ -264}{ 188}
\.{   48}{  34}
\.{   24}{  46}
\.{   20}{  48}
\.{   16}{  50}
\.{   12}{  52}
\.{    4}{  56}
\.{    0}{  58}
\.{   -8}{  62}
\.{  -12}{  64}
\.{  -16}{  66}
\.{  -24}{  70}
\.{  -36}{  76}
\.{  -40}{  78}
\.{  -42}{  79}
\.{  -44}{  80}
\.{  -48}{  82}
\.{  -54}{  85}
\.{  -56}{  86}
\.{  -60}{  88}
\.{  -64}{  90}
\.{  -72}{  94}
\.{  -88}{ 102}
\.{  -96}{ 106}
\.{ -108}{ 112}
\.{ -120}{ 118}
\.{ -128}{ 122}
\.{ -136}{ 126}
\.{ -168}{ 142}
\.{ -192}{ 154}
\.{ -240}{ 178}
\.{ -384}{ 250}
\.{ -432}{ 274}
\.{ -564}{ 340}
\.{   36}{  42}
\.{   32}{  44}
\.{   24}{  48}
\.{   16}{  52}
\.{   12}{  54}
\.{    0}{  60}
\.{  -12}{  66}
\.{  -20}{  70}
\.{  -24}{  72}
\.{  -32}{  76}
\.{  -36}{  78}
\.{  -56}{  88}
\.{  -60}{  90}
\.{  -72}{  96}
\.{  -84}{ 102}
\.{  -96}{ 108}
\.{ -108}{ 114}
\.{ -132}{ 126}
\.{ -168}{ 144}
\.{ -504}{ 312}
\.{   48}{  38}
\.{   36}{  44}
\.{   24}{  50}
\.{   16}{  54}
\.{   12}{  56}
\.{    0}{  62}
\.{   -6}{  65}
\.{   -8}{  66}
\.{  -18}{  71}
\.{  -24}{  74}
\.{  -32}{  78}
\.{  -48}{  86}
\.{  -72}{  98}
\.{  -80}{ 102}
\.{  -96}{ 110}
\.{ -120}{ 122}
\.{ -144}{ 134}
\.{ -228}{ 176}
\.{ -288}{ 206}
\.{ -336}{ 230}
\.{   48}{  40}
\.{   42}{  43}
\.{   24}{  52}
\.{   18}{  55}
\.{   16}{  56}
\.{   12}{  58}
\.{    8}{  60}
\.{    0}{  64}
\.{   -6}{  67}
\.{   -8}{  68}
\.{  -12}{  70}
\.{  -16}{  72}
\.{  -24}{  76}
\.{  -48}{  88}
\.{  -64}{  96}
\.{  -66}{  97}
\.{  -72}{ 100}
\.{  -96}{ 112}
\.{ -120}{ 124}
\.{ -168}{ 148}
\.{ -348}{ 238}
\.{ -396}{ 262}
\.{ -408}{ 268}
\.{ -540}{ 334}
\.{   48}{  42}
\.{   44}{  44}
\.{   32}{  50}
\.{   24}{  54}
\.{   18}{  57}
\.{    8}{  62}
\.{    0}{  66}
\.{   -8}{  70}
\.{  -12}{  72}
\.{  -16}{  74}
\.{  -24}{  78}
\.{  -36}{  84}
\.{  -40}{  86}
\.{  -48}{  90}
\.{  -72}{ 102}
\.{  -96}{ 114}
\.{ -108}{ 120}
\.{ -116}{ 124}
\.{ -144}{ 138}
\.{ -216}{ 174}
\.{ -480}{ 306}
\.{   48}{  44}
\.{   36}{  50}
\.{   24}{  56}
\.{   12}{  62}
\.{    6}{  65}
\.{    0}{  68}
\.{  -24}{  80}
\.{  -30}{  83}
\.{  -48}{  92}
\.{  -60}{  98}
\.{ -100}{ 118}
\.{ -120}{ 128}
\.{ -136}{ 136}
\.{ -168}{ 152}
\.{ -180}{ 158}
\.{ -240}{ 188}
\.{ -312}{ 224}
\.{   48}{  46}
\.{   42}{  49}
\.{   40}{  50}
\.{   36}{  52}
\.{   32}{  54}
\.{   24}{  58}
\.{   12}{  64}
\.{    8}{  66}
\.{    6}{  67}
\.{    0}{  70}
\.{  -12}{  76}
\.{  -16}{  78}
\.{  -24}{  82}
\.{  -36}{  88}
\.{  -40}{  90}
\.{  -48}{  94}
\.{  -96}{ 118}
\.{ -120}{ 130}
\.{ -160}{ 150}
\.{ -192}{ 166}
\.{ -288}{ 214}
\.{ -324}{ 232}
\.{ -528}{ 334}
\.{   60}{  42}
\.{   48}{  48}
\.{   36}{  54}
\.{   30}{  57}
\.{   24}{  60}
\.{   16}{  64}
\.{   12}{  66}
\.{    8}{  68}
\.{  -12}{  78}
\.{  -24}{  84}
\.{  -48}{  96}
\.{  -72}{ 108}
\.{  -96}{ 120}
\.{ -120}{ 132}
\.{ -216}{ 180}
\.{ -372}{ 258}
\.{   60}{  44}
\.{   48}{  50}
\.{   40}{  54}
\.{   24}{  62}
\.{   16}{  66}
\.{    8}{  70}
\.{    0}{  74}
\.{  -24}{  86}
\.{  -48}{  98}
\.{  -64}{ 106}
\.{  -96}{ 122}
\.{ -120}{ 134}
\.{ -192}{ 170}
\.{ -196}{ 172}
\.{ -216}{ 182}
\.{ -276}{ 212}
\.{   60}{  46}
\.{   54}{  49}
\.{   48}{  52}
\.{   36}{  58}
\.{   30}{  61}
\.{   24}{  64}
\.{   12}{  70}
\.{    0}{  76}
\.{  -12}{  82}
\.{  -24}{  88}
\.{  -36}{  94}
\.{  -48}{ 100}
\.{  -72}{ 112}
\.{ -120}{ 136}
\.{ -144}{ 148}
\.{ -156}{ 154}
\.{ -160}{ 156}
\.{ -252}{ 202}
\.{ -300}{ 226}
\.{ -372}{ 262}
\.{   72}{  42}
\.{   60}{  48}
\.{   48}{  54}
\.{   36}{  60}
\.{   30}{  63}
\.{   24}{  66}
\.{   12}{  72}
\.{    0}{  78}
\.{  -16}{  86}
\.{  -24}{  90}
\.{  -48}{ 102}
\.{  -80}{ 118}
\.{  -96}{ 126}
\.{ -120}{ 138}
\.{ -128}{ 142}
\.{ -144}{ 150}
\.{ -180}{ 168}
\.{ -288}{ 222}
\.{   72}{  44}
\.{   60}{  50}
\.{   32}{  64}
\.{   20}{  70}
\.{   18}{  71}
\.{   16}{  72}
\.{    0}{  80}
\.{  -12}{  86}
\.{  -32}{  96}
\.{  -48}{ 104}
\.{  -72}{ 116}
\.{ -252}{ 206}
\.{ -300}{ 230}
\.{   72}{  46}
\.{   64}{  50}
\.{   60}{  52}
\.{   48}{  58}
\.{   40}{  62}
\.{   24}{  70}
\.{   16}{  74}
\.{   12}{  76}
\.{    0}{  82}
\.{  -24}{  94}
\.{  -48}{ 106}
\.{  -60}{ 112}
\.{  -72}{ 118}
\.{ -120}{ 142}
\.{ -192}{ 178}
\.{ -228}{ 196}
\.{ -288}{ 226}
\.{   72}{  48}
\.{   48}{  60}
\.{   44}{  62}
\.{   12}{  78}
\.{  -12}{  90}
\.{  -36}{ 102}
\.{  -48}{ 108}
\.{  -72}{ 120}
\.{ -180}{ 174}
\.{ -216}{ 192}
\.{   80}{  46}
\.{   72}{  50}
\.{   64}{  54}
\.{   48}{  62}
\.{   36}{  68}
\.{   32}{  70}
\.{   16}{  78}
\.{    0}{  86}
\.{  -16}{  94}
\.{  -24}{  98}
\.{  -48}{ 110}
\.{  -56}{ 114}
\.{  -76}{ 124}
\.{  -96}{ 134}
\.{ -176}{ 174}
\.{ -240}{ 206}
\.{   72}{  52}
\.{   60}{  58}
\.{   48}{  64}
\.{   40}{  68}
\.{   36}{  70}
\.{   32}{  72}
\.{   24}{  76}
\.{   12}{  82}
\.{    0}{  88}
\.{  -12}{  94}
\.{  -24}{ 100}
\.{  -32}{ 104}
\.{  -36}{ 106}
\.{  -72}{ 124}
\.{  -84}{ 130}
\.{ -120}{ 148}
\.{ -140}{ 158}
\.{ -192}{ 184}
\.{   84}{  48}
\.{   72}{  54}
\.{   60}{  60}
\.{   56}{  62}
\.{   52}{  64}
\.{   48}{  66}
\.{   24}{  78}
\.{    0}{  90}
\.{  -24}{ 102}
\.{  -36}{ 108}
\.{  -40}{ 110}
\.{  -52}{ 116}
\.{  -96}{ 138}
\.{ -168}{ 174}
\.{   72}{  56}
\.{   60}{  62}
\.{   48}{  68}
\.{   36}{  74}
\.{   32}{  76}
\.{   24}{  80}
\.{   12}{  86}
\.{    0}{  92}
\.{  -72}{ 128}
\.{ -120}{ 152}
\.{ -144}{ 164}
\.{   80}{  54}
\.{   72}{  58}
\.{   64}{  62}
\.{   52}{  68}
\.{   48}{  70}
\.{   36}{  76}
\.{   32}{  78}
\.{   24}{  82}
\.{   16}{  86}
\.{    0}{  94}
\.{  -12}{ 100}
\.{  -36}{ 112}
\.{  -48}{ 118}
\.{  -60}{ 124}
\.{  -96}{ 142}
\.{ -168}{ 178}
\.{ -180}{ 184}
\.{ -192}{ 190}
\.{ -480}{ 334}
\.{   84}{  54}
\.{   72}{  60}
\.{   56}{  68}
\.{   48}{  72}
\.{   36}{  78}
\.{   24}{  84}
\.{   12}{  90}
\.{  -12}{ 102}
\.{  -40}{ 116}
\.{  -72}{ 132}
\.{  -84}{ 138}
\.{  -96}{ 144}
\.{ -120}{ 156}
\.{ -264}{ 228}
\.{ -276}{ 234}
\.{ -420}{ 306}
\.{   88}{  54}
\.{   80}{  58}
\.{   72}{  62}
\.{   64}{  66}
\.{   60}{  68}
\.{   54}{  71}
\.{   48}{  74}
\.{   40}{  78}
\.{   30}{  83}
\.{   24}{  86}
\.{    0}{  98}
\.{  -20}{ 108}
\.{  -40}{ 118}
\.{  -48}{ 122}
\.{  -56}{ 126}
\.{ -120}{ 158}
\.{ -160}{ 178}
\.{   84}{  58}
\.{   80}{  60}
\.{   78}{  61}
\.{   72}{  64}
\.{   64}{  68}
\.{   48}{  76}
\.{   42}{  79}
\.{   24}{  88}
\.{  -24}{ 112}
\.{  -48}{ 124}
\.{ -144}{ 172}
\.{ -168}{ 184}
\.{ -324}{ 262}
\.{   96}{  54}
\.{   80}{  62}
\.{   72}{  66}
\.{   66}{  69}
\.{   56}{  74}
\.{   48}{  78}
\.{   44}{  80}
\.{   36}{  84}
\.{   24}{  90}
\.{   16}{  94}
\.{  -12}{ 108}
\.{  -24}{ 114}
\.{  -32}{ 118}
\.{ -144}{ 174}
\.{ -192}{ 198}
\.{   96}{  56}
\.{   84}{  62}
\.{   72}{  68}
\.{   60}{  74}
\.{   48}{  80}
\.{    0}{ 104}
\.{  -24}{ 116}
\.{  -72}{ 140}
\.{ -216}{ 212}
\.{   96}{  58}
\.{   84}{  64}
\.{   72}{  70}
\.{   60}{  76}
\.{   48}{  82}
\.{   40}{  86}
\.{   36}{  88}
\.{   24}{  94}
\.{   12}{ 100}
\.{    0}{ 106}
\.{  -72}{ 142}
\.{  -96}{ 154}
\.{ -240}{ 226}
\.{  108}{  54}
\.{   96}{  60}
\.{   84}{  66}
\.{   80}{  68}
\.{   64}{  76}
\.{   60}{  78}
\.{   32}{  92}
\.{   12}{ 102}
\.{  -16}{ 116}
\.{  -24}{ 120}
\.{  -36}{ 126}
\.{  -60}{ 138}
\.{  -72}{ 144}
\.{ -120}{ 168}
\.{   96}{  62}
\.{   80}{  70}
\.{   72}{  74}
\.{   64}{  78}
\.{   48}{  86}
\.{   40}{  90}
\.{   24}{  98}
\.{    0}{ 110}
\.{  -12}{ 116}
\.{  -40}{ 130}
\.{  -48}{ 134}
\.{  -96}{ 158}
\.{ -144}{ 182}
\.{ -192}{ 206}
\.{  102}{  61}
\.{   90}{  67}
\.{   88}{  68}
\.{   80}{  72}
\.{   72}{  76}
\.{   64}{  80}
\.{   60}{  82}
\.{   54}{  85}
\.{   48}{  88}
\.{   36}{  94}
\.{   32}{  96}
\.{   24}{ 100}
\.{  -36}{ 130}
\.{  -72}{ 148}
\.{ -108}{ 166}
\.{ -132}{ 178}
\.{ -240}{ 232}
\.{   96}{  66}
\.{   88}{  70}
\.{   80}{  74}
\.{   72}{  78}
\.{   60}{  84}
\.{   56}{  86}
\.{   48}{  90}
\.{   24}{ 102}
\.{   12}{ 108}
\.{  -48}{ 138}
\.{  -60}{ 144}
\.{   80}{  76}
\.{   72}{  80}
\.{   48}{  92}
\.{  -36}{ 134}
\.{  102}{  67}
\.{   96}{  70}
\.{   78}{  79}
\.{   72}{  82}
\.{   64}{  86}
\.{   60}{  88}
\.{   48}{  94}
\.{   20}{ 108}
\.{    0}{ 118}
\.{  -24}{ 130}
\.{  -48}{ 142}
\.{  -96}{ 166}
\.{ -144}{ 190}
\.{ -432}{ 334}
\.{  108}{  66}
\.{  104}{  68}
\.{   96}{  72}
\.{   84}{  78}
\.{   72}{  84}
\.{   66}{  87}
\.{   60}{  90}
\.{   48}{  96}
\.{   36}{ 102}
\.{   32}{ 104}
\.{  -24}{ 132}
\.{ -372}{ 306}
\.{  120}{  62}
\.{  108}{  68}
\.{   96}{  74}
\.{   72}{  86}
\.{   64}{  90}
\.{   48}{  98}
\.{   12}{ 116}
\.{    0}{ 122}
\.{  -12}{ 128}
\.{  -24}{ 134}
\.{  -48}{ 146}
\.{  -72}{ 158}
\.{  -84}{ 164}
\.{  120}{  64}
\.{  108}{  70}
\.{   96}{  76}
\.{   92}{  78}
\.{   84}{  82}
\.{   72}{  88}
\.{   48}{ 100}
\.{   36}{ 106}
\.{   24}{ 112}
\.{   16}{ 116}
\.{  -12}{ 130}
\.{  -84}{ 166}
\.{ -108}{ 178}
\.{ -420}{ 334}
\.{  120}{  66}
\.{  112}{  70}
\.{  108}{  72}
\.{  104}{  74}
\.{   96}{  78}
\.{   84}{  84}
\.{   48}{ 102}
\.{   36}{ 108}
\.{   24}{ 114}
\.{    0}{ 126}
\.{ -288}{ 270}
\.{  120}{  68}
\.{  108}{  74}
\.{   96}{  80}
\.{   72}{  92}
\.{   60}{  98}
\.{   48}{ 104}
\.{   24}{ 116}
\.{  -72}{ 164}
\.{  120}{  70}
\.{  108}{  76}
\.{  100}{  80}
\.{   96}{  82}
\.{   84}{  88}
\.{   72}{  94}
\.{   66}{  97}
\.{   48}{ 106}
\.{   40}{ 110}
\.{   36}{ 112}
\.{    0}{ 130}
\.{  -24}{ 142}
\.{  -84}{ 172}
\.{ -120}{ 190}
\.{ -192}{ 226}
\.{ -264}{ 262}
\.{  126}{  69}
\.{  120}{  72}
\.{  112}{  76}
\.{  108}{  78}
\.{  104}{  80}
\.{  102}{  81}
\.{   84}{  90}
\.{   76}{  94}
\.{   72}{  96}
\.{   48}{ 108}
\.{   24}{ 120}
\.{  -84}{ 174}
\.{ -156}{ 210}
\.{ -216}{ 240}
\.{  120}{  74}
\.{  112}{  78}
\.{  108}{  80}
\.{  102}{  83}
\.{   96}{  86}
\.{   72}{  98}
\.{   48}{ 110}
\.{   32}{ 118}
\.{   12}{ 128}
\.{  -48}{ 158}
\.{  -64}{ 166}
\.{ -144}{ 206}
\.{  120}{  76}
\.{   96}{  88}
\.{   84}{  94}
\.{   80}{  96}
\.{   72}{ 100}
\.{   40}{ 116}
\.{   12}{ 130}
\.{  -24}{ 148}
\.{  -60}{ 166}
\.{ -108}{ 190}
\.{ -120}{ 196}
\.{ -132}{ 202}
\.{ -180}{ 226}
\.{ -216}{ 244}
\.{  132}{  72}
\.{  128}{  74}
\.{  120}{  78}
\.{  116}{  80}
\.{  108}{  84}
\.{  104}{  86}
\.{   96}{  90}
\.{   90}{  93}
\.{   72}{ 102}
\.{   64}{ 106}
\.{   40}{ 118}
\.{    0}{ 138}
\.{  -36}{ 156}
\.{  -56}{ 166}
\.{  -96}{ 186}
\.{  132}{  74}
\.{  120}{  80}
\.{  108}{  86}
\.{   96}{  92}
\.{   84}{  98}
\.{  -12}{ 146}
\.{  132}{  76}
\.{  126}{  79}
\.{  120}{  82}
\.{  112}{  86}
\.{  108}{  88}
\.{   96}{  94}
\.{   80}{ 102}
\.{   60}{ 112}
\.{   56}{ 114}
\.{   52}{ 116}
\.{   48}{ 118}
\.{   24}{ 130}
\.{    0}{ 142}
\.{  -48}{ 166}
\.{  -96}{ 190}
\.{ -144}{ 214}
\.{ -396}{ 340}
\.{  132}{  78}
\.{  120}{  84}
\.{  108}{  90}
\.{   72}{ 108}
\.{   36}{ 126}
\.{   24}{ 132}
\.{ -336}{ 312}
\.{  144}{  74}
\.{  136}{  78}
\.{  132}{  80}
\.{  120}{  86}
\.{   96}{  98}
\.{   48}{ 122}
\.{   24}{ 134}
\.{  144}{  76}
\.{  120}{  88}
\.{  112}{  92}
\.{  108}{  94}
\.{   48}{ 124}
\.{   36}{ 130}
\.{  -24}{ 160}
\.{  -60}{ 178}
\.{  -84}{ 190}
\.{ -240}{ 268}
\.{  144}{  78}
\.{  128}{  86}
\.{  126}{  87}
\.{  120}{  90}
\.{  108}{  96}
\.{   96}{ 102}
\.{   40}{ 130}
\.{    0}{ 150}
\.{  -12}{ 156}
\.{  120}{  92}
\.{  112}{  96}
\.{   72}{ 116}
\.{   36}{ 134}
\.{   12}{ 146}
\.{ -120}{ 212}
\.{  144}{  82}
\.{  138}{  85}
\.{  136}{  86}
\.{  120}{  94}
\.{   96}{ 106}
\.{   72}{ 118}
\.{   56}{ 126}
\.{  -36}{ 172}
\.{ -156}{ 232}
\.{  128}{  92}
\.{  120}{  96}
\.{   96}{ 108}
\.{   72}{ 120}
\.{  -96}{ 204}
\.{  144}{  86}
\.{  140}{  88}
\.{  132}{  92}
\.{  120}{  98}
\.{   96}{ 110}
\.{   80}{ 118}
\.{   48}{ 134}
\.{    0}{ 158}
\.{ -108}{ 212}
\.{  144}{  88}
\.{  120}{ 100}
\.{   96}{ 112}
\.{   72}{ 124}
\.{   24}{ 148}
\.{ -372}{ 346}
\.{  144}{  90}
\.{  120}{ 102}
\.{   96}{ 114}
\.{   76}{ 124}
\.{   72}{ 126}
\.{   48}{ 138}
\.{   12}{ 156}
\.{    0}{ 162}
\.{ -312}{ 318}
\.{  144}{  92}
\.{  120}{ 104}
\.{  112}{ 108}
\.{   72}{ 128}
\.{  -96}{ 212}
\.{  160}{  86}
\.{  156}{  88}
\.{  148}{  92}
\.{  144}{  94}
\.{  138}{  97}
\.{  132}{ 100}
\.{  128}{ 102}
\.{  120}{ 106}
\.{  112}{ 110}
\.{  108}{ 112}
\.{   96}{ 118}
\.{   48}{ 142}
\.{    0}{ 166}
\.{  -60}{ 196}
\.{  168}{  84}
\.{  150}{  93}
\.{  144}{  96}
\.{  132}{ 102}
\.{  120}{ 108}
\.{  108}{ 114}
\.{   72}{ 132}
\.{   60}{ 138}
\.{  160}{  90}
\.{  144}{  98}
\.{  132}{ 104}
\.{  120}{ 110}
\.{   96}{ 122}
\.{   48}{ 146}
\.{  168}{  88}
\.{  156}{  94}
\.{  132}{ 106}
\.{  128}{ 108}
\.{  120}{ 112}
\.{   84}{ 130}
\.{   24}{ 160}
\.{  -24}{ 184}
\.{ -132}{ 238}
\.{ -168}{ 256}
\.{  144}{ 102}
\.{  128}{ 110}
\.{  108}{ 120}
\.{   96}{ 126}
\.{   60}{ 144}
\.{   36}{ 156}
\.{  -32}{ 190}
\.{  144}{ 104}
\.{  128}{ 112}
\.{  120}{ 116}
\.{   72}{ 140}
\.{  -72}{ 212}
\.{  -84}{ 218}
\.{  168}{  94}
\.{  160}{  98}
\.{  152}{ 102}
\.{  144}{ 106}
\.{  120}{ 118}
\.{   72}{ 142}
\.{    0}{ 178}
\.{  -48}{ 202}
\.{  180}{  90}
\.{  168}{  96}
\.{  144}{ 108}
\.{   84}{ 138}
\.{   72}{ 144}
\.{  168}{  98}
\.{  160}{ 102}
\.{  144}{ 110}
\.{  120}{ 122}
\.{  116}{ 124}
\.{   96}{ 134}
\.{   48}{ 158}
\.{    0}{ 182}
\.{  168}{ 100}
\.{  156}{ 106}
\.{  152}{ 108}
\.{  144}{ 112}
\.{  120}{ 124}
\.{   72}{ 148}
\.{  -36}{ 202}
\.{ -216}{ 292}
\.{  156}{ 108}
\.{  152}{ 110}
\.{  128}{ 122}
\.{   96}{ 138}
\.{  -56}{ 214}
\.{  180}{  98}
\.{  160}{ 108}
\.{  156}{ 110}
\.{  120}{ 128}
\.{  -60}{ 218}
\.{ -120}{ 248}
\.{  176}{ 102}
\.{  168}{ 106}
\.{  160}{ 110}
\.{  156}{ 112}
\.{  144}{ 118}
\.{  120}{ 130}
\.{   96}{ 142}
\.{   48}{ 166}
\.{   36}{ 172}
\.{    0}{ 190}
\.{ -336}{ 358}
\.{  168}{ 108}
\.{  156}{ 114}
\.{  132}{ 126}
\.{  120}{ 132}
\.{   96}{ 144}
\.{ -276}{ 330}
\.{  184}{ 102}
\.{  176}{ 106}
\.{  168}{ 110}
\.{  144}{ 122}
\.{  136}{ 126}
\.{  120}{ 134}
\.{   72}{ 158}
\.{   56}{ 166}
\.{    0}{ 194}
\.{  -36}{ 212}
\.{  180}{ 106}
\.{  168}{ 112}
\.{  144}{ 124}
\.{  120}{ 136}
\.{   60}{ 166}
\.{  -36}{ 214}
\.{ -180}{ 286}
\.{  192}{ 102}
\.{  168}{ 114}
\.{  156}{ 120}
\.{  120}{ 138}
\.{   64}{ 166}
\.{  168}{ 116}
\.{  160}{ 120}
\.{   72}{ 164}
\.{  200}{ 102}
\.{  192}{ 106}
\.{  184}{ 110}
\.{  180}{ 112}
\.{  168}{ 118}
\.{  156}{ 124}
\.{   96}{ 154}
\.{  -96}{ 250}
\.{  196}{ 106}
\.{  184}{ 112}
\.{  168}{ 120}
\.{  156}{ 126}
\.{  136}{ 136}
\.{  -36}{ 222}
\.{  204}{ 104}
\.{  200}{ 106}
\.{  192}{ 110}
\.{  168}{ 122}
\.{  152}{ 130}
\.{  144}{ 134}
\.{  128}{ 142}
\.{   96}{ 158}
\.{   84}{ 164}
\.{   32}{ 190}
\.{    0}{ 206}
\.{  -48}{ 230}
\.{  192}{ 112}
\.{  180}{ 118}
\.{  168}{ 124}
\.{  120}{ 148}
\.{   84}{ 166}
\.{   60}{ 178}
\.{  204}{ 108}
\.{  192}{ 114}
\.{  180}{ 120}
\.{  156}{ 132}
\.{  144}{ 138}
\.{  208}{ 108}
\.{  192}{ 116}
\.{  168}{ 128}
\.{  120}{ 152}
\.{  192}{ 118}
\.{   96}{ 166}
\.{   84}{ 172}
\.{    0}{ 214}
\.{  120}{ 156}
\.{   84}{ 174}
\.{ -120}{ 276}
\.{  192}{ 122}
\.{  168}{ 134}
\.{  120}{ 158}
\.{ -120}{ 278}
\.{  192}{ 124}
\.{  144}{ 148}
\.{  108}{ 166}
\.{   36}{ 202}
\.{  204}{ 120}
\.{  144}{ 150}
\.{  216}{ 116}
\.{  192}{ 128}
\.{  168}{ 140}
\.{  184}{ 134}
\.{  168}{ 142}
\.{   60}{ 196}
\.{   48}{ 202}
\.{  216}{ 120}
\.{  204}{ 126}
\.{  180}{ 138}
\.{  168}{ 144}
\.{  140}{ 158}
\.{  120}{ 168}
\.{  192}{ 134}
\.{  160}{ 150}
\.{   36}{ 212}
\.{ -288}{ 374}
\.{  216}{ 124}
\.{  204}{ 130}
\.{  168}{ 148}
\.{  156}{ 154}
\.{  108}{ 178}
\.{   84}{ 190}
\.{   36}{ 214}
\.{ -228}{ 346}
\.{  216}{ 126}
\.{  200}{ 134}
\.{  192}{ 138}
\.{   96}{ 186}
\.{  216}{ 128}
\.{  204}{ 134}
\.{  168}{ 152}
\.{  160}{ 156}
\.{  144}{ 164}
\.{  192}{ 142}
\.{   96}{ 190}
\.{    0}{ 238}
\.{  -96}{ 286}
\.{  228}{ 126}
\.{  216}{ 132}
\.{   36}{ 222}
\.{  232}{ 126}
\.{  216}{ 134}
\.{   56}{ 214}
\.{    0}{ 242}
\.{  -48}{ 266}
\.{  240}{ 124}
\.{  228}{ 130}
\.{  144}{ 172}
\.{  132}{ 178}
\.{  108}{ 190}
\.{  240}{ 126}
\.{  228}{ 132}
\.{  192}{ 150}
\.{  144}{ 174}
\.{    0}{ 246}
\.{  180}{ 158}
\.{   72}{ 212}
\.{   60}{ 218}
\.{  232}{ 134}
\.{  216}{ 142}
\.{  200}{ 150}
\.{  192}{ 154}
\.{  120}{ 190}
\.{  228}{ 138}
\.{  216}{ 144}
\.{   96}{ 204}
\.{  240}{ 134}
\.{  228}{ 140}
\.{  224}{ 142}
\.{  144}{ 182}
\.{   48}{ 230}
\.{  252}{ 130}
\.{  120}{ 196}
\.{  240}{ 138}
\.{  216}{ 150}
\.{  180}{ 168}
\.{  168}{ 174}
\.{  228}{ 146}
\.{  216}{ 152}
\.{   96}{ 212}
\.{   84}{ 218}
\.{  240}{ 142}
\.{  192}{ 166}
\.{  176}{ 174}
\.{  168}{ 178}
\.{  144}{ 190}
\.{    0}{ 262}
\.{  260}{ 134}
\.{  240}{ 144}
\.{  180}{ 174}
\.{  216}{ 158}
\.{  192}{ 170}
\.{  108}{ 212}
\.{  252}{ 142}
\.{  168}{ 184}
\.{  132}{ 202}
\.{  240}{ 150}
\.{  196}{ 172}
\.{  216}{ 164}
\.{  120}{ 212}
\.{  232}{ 158}
\.{  192}{ 178}
\.{  180}{ 184}
\.{ -240}{ 394}
\.{  264}{ 144}
\.{  252}{ 150}
\.{ -180}{ 366}
\.{  240}{ 158}
\.{  144}{ 206}
\.{  264}{ 148}
\.{  260}{ 150}
\.{  236}{ 162}
\.{  192}{ 184}
\.{  -84}{ 322}
\.{  240}{ 162}
\.{  216}{ 174}
\.{  228}{ 170}
\.{  272}{ 150}
\.{  240}{ 166}
\.{  192}{ 190}
\.{  144}{ 214}
\.{    0}{ 286}
\.{  280}{ 148}
\.{  276}{ 150}
\.{  252}{ 162}
\.{  216}{ 180}
\.{  156}{ 210}
\.{  288}{ 146}
\.{  264}{ 158}
\.{  228}{ 176}
\.{  216}{ 182}
\.{   48}{ 266}
\.{  276}{ 154}
\.{  204}{ 190}
\.{  264}{ 162}
\.{  256}{ 166}
\.{  240}{ 174}
\.{  192}{ 198}
\.{  288}{ 152}
\.{  264}{ 164}
\.{  296}{ 150}
\.{  240}{ 178}
\.{   96}{ 250}
\.{  216}{ 192}
}
\.{  288}{ 158}
\.{  256}{ 174}
\.{  192}{ 206}
\.{    0}{ 302}
\.{  264}{ 172}
\.{  252}{ 178}
\.{  132}{ 238}
\.{  288}{ 162}
\.{  240}{ 188}
\.{  120}{ 248}
\.{  288}{ 166}
\.{  276}{ 172}
\.{  228}{ 196}
\.{  156}{ 232}
\.{  276}{ 174}
\.{  296}{ 166}
\.{  288}{ 170}
\.{  272}{ 178}
\.{  264}{ 184}
\.{  300}{ 168}
\.{  264}{ 188}
\.{  216}{ 212}
\.{  312}{ 166}
\.{  192}{ 226}
\.{  288}{ 182}
\.{  240}{ 206}
\.{  312}{ 172}
\.{  304}{ 176}
\.{  252}{ 202}
\.{  -60}{ 358}
\.{  324}{ 168}
\.{  300}{ 180}
\.{  288}{ 186}
\.{  276}{ 192}
\.{  288}{ 188}
\.{  252}{ 206}
\.{  312}{ 178}
\.{  288}{ 190}
\.{   96}{ 286}
\.{  120}{ 276}
\.{  120}{ 278}
\.{  168}{ 256}
\.{  336}{ 178}
\.{  324}{ 184}
\.{  312}{ 190}
\.{  240}{ 226}
\.{  312}{ 192}
\.{  300}{ 198}
\.{  216}{ 240}
\.{  288}{ 206}
\.{  312}{ 196}
\.{  216}{ 244}
\.{ -156}{ 430}
\.{  336}{ 188}
\.{  288}{ 214}
\.{    0}{ 358}
\.{  348}{ 186}
\.{  264}{ 228}
\.{  336}{ 194}
\.{   84}{ 322}
\.{  336}{ 198}
\.{  288}{ 222}
\.{  340}{ 198}
\.{  360}{ 190}
\.{  336}{ 202}
\.{  288}{ 226}
\.{  348}{ 198}
\.{  276}{ 234}
\.{  336}{ 206}
\.{  300}{ 226}
\.{  180}{ 286}
\.{  372}{ 194}
\.{  312}{ 224}
\.{  300}{ 230}
\.{  372}{ 202}
\.{  368}{ 204}
\.{   60}{ 358}
\.{  360}{ 212}
\.{  324}{ 232}
\.{  264}{ 262}
\.{  384}{ 204}
\.{  336}{ 230}
\.{  216}{ 292}
\.{  376}{ 214}
\.{  360}{ 228}
\.{  384}{ 218}
\.{  372}{ 226}
\.{  348}{ 238}
\.{  288}{ 270}
\.{  408}{ 212}
\.{  396}{ 222}
\.{  324}{ 262}
\.{  420}{ 218}
\.{  408}{ 224}
\.{  384}{ 242}
\.{  408}{ 232}
\.{  420}{ 230}
\.{  384}{ 250}
\.{  408}{ 240}
\.{  372}{ 258}
\.{  -60}{ 474}
\.{    0}{ 446}
\.{  372}{ 262}
\.{  432}{ 238}
\.{  180}{ 366}
\.{  432}{ 242}
\.{  396}{ 262}
\.{  228}{ 346}
\.{  456}{ 234}
\.{  276}{ 330}
\.{  408}{ 268}
\.{  312}{ 318}
\.{  456}{ 248}
\.{  432}{ 266}
\.{  456}{ 256}
\.{  480}{ 246}
\.{  456}{ 264}
\.{  456}{ 272}
\.{  492}{ 256}
\.{  480}{ 262}
\.{    0}{ 502}
\.{   60}{ 474}
\.{  156}{ 430}
\.{  240}{ 394}
\.{  420}{ 306}
\.{  288}{ 374}
\.{  468}{ 286}
\.{  480}{ 286}
\.{  336}{ 358}
\.{  372}{ 346}
\.{  396}{ 340}
\.{  528}{ 278}
\.{  540}{ 274}
\.{  528}{ 286}
\.{  432}{ 334}
\.{  516}{ 302}
\.{  480}{ 334}
\.{  552}{ 306}
\.{  528}{ 334}
\.{  564}{ 322}
\.{  540}{ 334}
\.{  564}{ 330}
\.{  564}{ 340}
\.{  612}{ 330}
\.{  588}{ 346}
\.{  624}{ 330}
\.{  624}{ 358}
\.{  660}{ 366}
\.{  672}{ 374}
\.{  732}{ 386}
\.{  720}{ 394}
\.{  804}{ 430}
\.{  900}{ 474}
\.{  960}{ 502}
\begin{center}
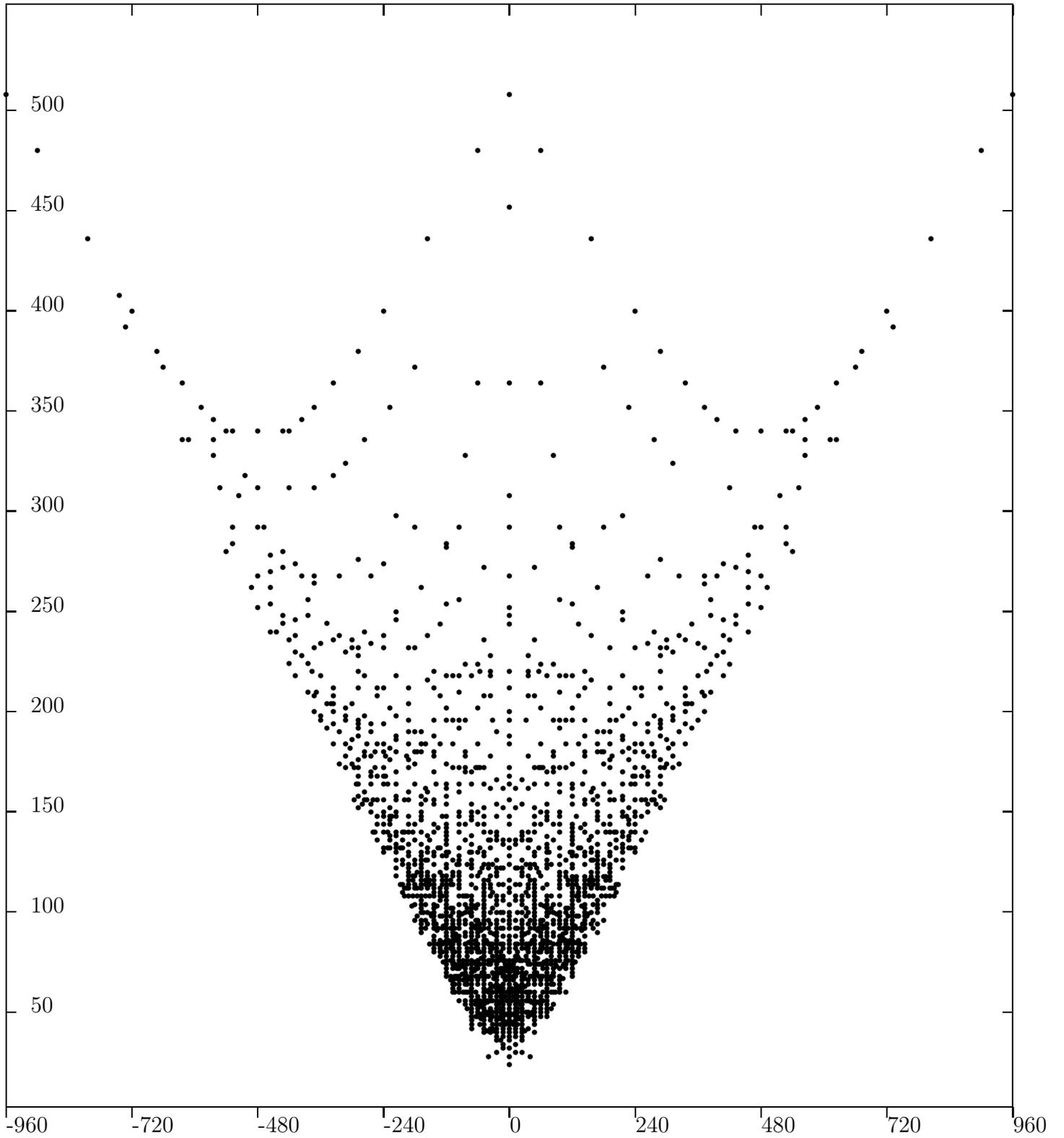
\parbox{6.4truein}{\noindent {\bf Fig. 1}~~{\it A plot of Euler numbers against
 ${\bar n}_g+n_g$ for the 1898 spectra of all the LG potentials and phase
orbifolds constructed.}}
\end{center}
\end{center}

\begin{center}
\plot{3.3truein}{\tiny{$\bullet$}}
\nobreak
\frame
\nobreak
\.{ -24}{ 12}
\.{ -64}{ 32}
\.{ -72}{ 36}
\.{-168}{ 84}
\.{ -40}{ 22}
\.{ -72}{ 38}
\.{ -96}{ 50}
\.{-108}{ 56}
\.{-120}{ 62}
\.{-144}{ 74}
\.{ -60}{ 34}
\.{-108}{ 58}
\.{-120}{ 64}
\.{ -48}{ 30}
\.{ -64}{ 38}
\.{ -84}{ 48}
\.{ -96}{ 54}
\.{-112}{ 62}
\.{-120}{ 66}
\.{-160}{ 86}
\.{ -18}{ 17}
\.{ -36}{ 26}
\.{ -72}{ 44}
\.{ -96}{ 56}
\.{-132}{ 74}
\.{ -40}{ 30}
\.{ -56}{ 38}
\.{ -72}{ 46}
\.{ -80}{ 50}
\.{ -88}{ 54}
\.{ -96}{ 58}
\.{-104}{ 62}
\.{ -24}{ 24}
\.{ -60}{ 42}
\.{-120}{ 72}
\.{ -48}{ 38}
\.{ -72}{ 50}
\.{-144}{ 86}
\.{ -32}{ 32}
\.{ -36}{ 34}
\.{ -60}{ 46}
\.{ -64}{ 48}
\.{ -72}{ 52}
\.{ -80}{ 56}
\.{ -88}{ 60}
\.{-108}{ 70}
\.{-120}{ 76}
\.{-180}{106}
\.{   0}{ 18}
\.{ -36}{ 36}
\.{ -48}{ 42}
\.{ -96}{ 66}
\.{-108}{ 72}
\.{-120}{ 78}
\.{-136}{ 86}
\.{ -72}{ 56}
\.{   0}{ 22}
\.{ -48}{ 46}
\.{ -60}{ 52}
\.{ -96}{ 70}
\.{ -24}{ 36}
\.{ -36}{ 42}
\.{ -24}{ 38}
\.{ -36}{ 44}
\.{ -72}{ 62}
\.{ -24}{ 40}
\.{-120}{ 88}
\.{   0}{ 30}
\.{ -24}{ 44}
\.{ -72}{ 68}
\.{  -8}{ 38}
\.{ -24}{ 46}
\.{ -40}{ 54}
\.{ -48}{ 58}
\.{ -24}{ 48}
\.{   0}{ 38}
\.{ -12}{ 46}
\.{   0}{ 42}
\.{  -8}{ 46}
\.{ -48}{ 66}
\.{   0}{ 46}
\.{ -24}{ 58}
\.{ -48}{ 70}
\.{ -24}{ 60}
\.{   0}{ 50}
\.{  24}{ 40}
\.{   8}{ 48}
\.{  24}{ 42}
\.{  24}{ 46}
\.{ -24}{ 70}
\.{   0}{ 62}
\.{  24}{ 52}
\.{  48}{ 46}
\.{  48}{ 54}
\.{  48}{ 58}
\.{  72}{ 62}
\begin{center}
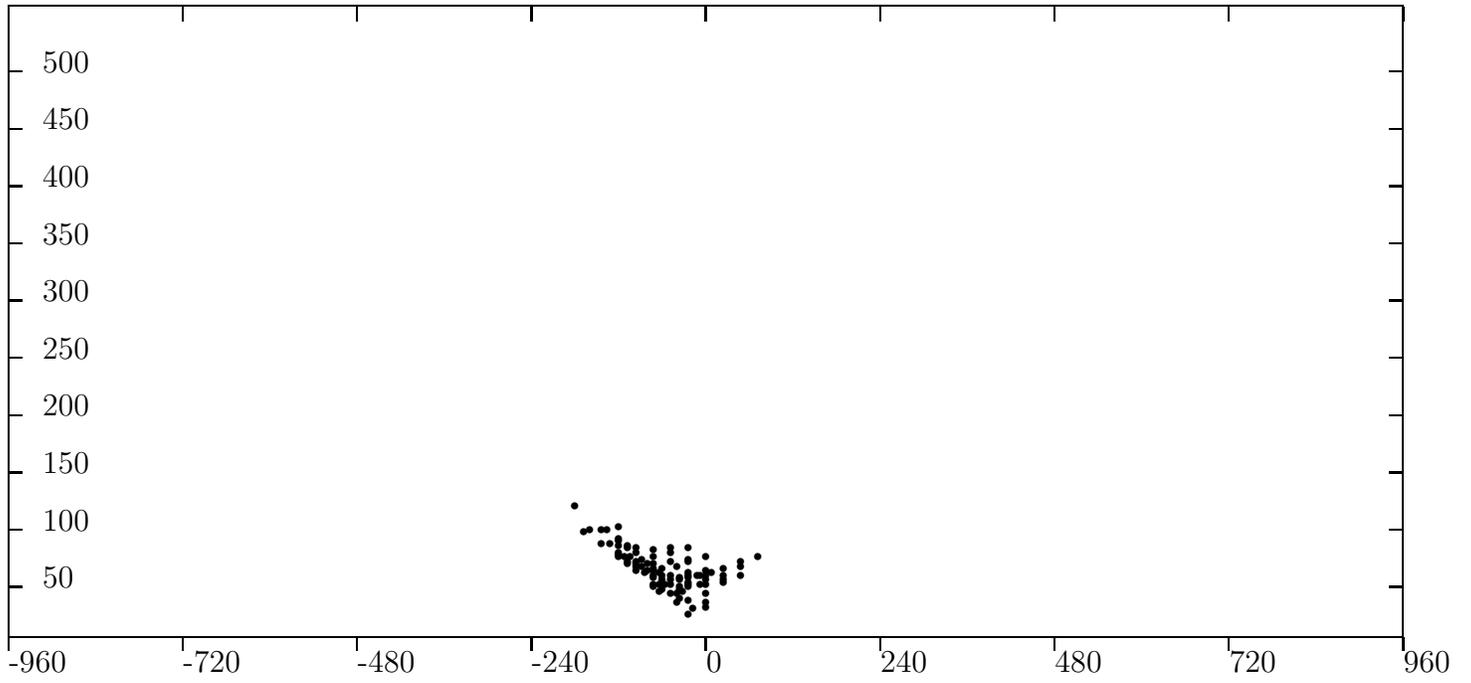
\parbox{6.4truein}{\noindent {\bf Fig. 2}~~{\it A plot of Euler numbers against
the total number of particles for the orbifolds with respect to cyclic
permutations.}}
\end{center}
\end{center}

\end{document}